\documentclass{aa}  
\usepackage{natbib}
\bibpunct{(}{)}{;}{a}{}{,}
\usepackage{graphicx}
\usepackage{rotating}
\usepackage{txfonts}
\def\feiil{[Fe~{\sc ii}]~1.64$\mu$m}
\def\feii{[Fe~{\sc ii}]}
\def\oi{[O~{\sc i}]}
\def\sii{[S~{\sc ii}]}
\def\h2{H$_2~$2.12$\mu$m}

\def\mic{$\mu$m}
\def\kms{km~s$^{-1}$}
\def\cm{cm$^{-3}$}
\def\ne{$n_{\rm e}$}
\def\xe{$x_{\rm e}$}
\def\Te{$T_{\rm e}$}
\def\Lsun{L$_\odot$}
\def\Msun{M$_\odot$}
\def\yr{yr$^{-1}$}
\def\arcs{$^{\prime\prime}$}
\def\macc{${\dot M}_{\rm acc}$}

\begin{document}

\title{Sub-arcsecond  \feii\ spectro-imaging of the DG~Tau jet} 
\subtitle{Periodic bubbles and a dusty disk wind ?} 
\author{
V. Agra-Amboage
          \inst{1}
          \and
          C. Dougados\inst{1}
          \and
          S. Cabrit\inst{2}
          \and
          J. Reunanen\inst{3} 
}
\institute{UJF-Grenoble 1/CNRS-INSU, Institut de Plan\'etologie et d'Astrophysique de Grenoble (IPAG) UMR 5274, Grenoble, F-38041, France 
              \email{vaa@fe.up.pt}
            \and
             LERMA, Observatoire de Paris, UMR 8112 du
             CNRS, 61 Avenue de l'Observatoire, 75014 Paris, France
           \email{sylvie.cabrit@obspm.fr}
           \and
           Tuorla Observatory, Department of Physics and Astronomy, University of 
           Turku, Vaisalantie 20, 21500 Piikkio, Finland
}
\offprints{V. Agra-Amboage}
\date{Received 07 October 2010; Accepted 03 May 2011}

\abstract
{The origin of protostellar jets as well as their impact on the regulation
of angular momentum and the inner disk physics are still crucial open
questions in star formation. } 
{We aim to test the different proposed ejection processes in T Tauri
stars through high-angular resolution observations of forbidden-line
emission from the inner DG~Tauri microjet.} 
{We present spectro-imaging observations of the DG~Tauri jet obtained
with SINFONI/VLT in the lines of \feii$\lambda$1.64 $\mu$m, 1.53$\mu$m
with 0\farcs15 angular resolution and R=3000 spectral resolution. We
analyze the morphology and kinematics, derive electronic densities and
mass-flux rates and discuss the implications for proposed jet
launching models.}
{ (1) We observe an onion-like velocity structure in \feii\ in the
blueshifted jet, similar to that observed in optical lines.
High-velocity (HV) gas  at $\simeq$ -200  \kms\ is collimated
inside a half-opening angle of 4\degr\ and medium-velocity (MV) gas
 at $\simeq$ -100 \kms  in a cone with an half-opening
angle 14\degr.  (2) Two  new  axial jet knots are detected in
the blue jet, as well as a more distant bubble with corresponding
counter-bubble. The periodic knot ejection timescale is revised
downward to 2.5 yrs.  (3) The redshifted jet is detected  only beyond
0\farcs7 from the star, yielding revised constraints on the disk
 surface density.  (4) From comparison to \oi\ data we infer iron depletion
of a factor 3 at high velocities and a factor 10 at speeds
below -100 \kms. (5)  The mass-fluxes in each of the medium and 
high-velocity components of the blueshifted lobe are $\simeq$ 1.6 $\pm$
0.8 $\times$ 10$^{-8}$ \Msun~yr$^{-1}$,  representing 
0.02--0.2 of the disk accretion rate.} 
{The medium-velocity conical \feii\ flow in the DG Tau jet is too fast
 and too narrow  to trace photo-evaporated matter from the disk atmosphere. 
Both its kinematics and collimation
cannot be reproduced by the X-wind, nor can the "conical
magnetospheric wind". The level
of  Fe gas phase depletion in the DG Tau medium-velocity component also rules out a stellar wind and a cocoon ejected
sideways from the high-velocity beam.  A quasi-steady centrifugal MHD
disk wind ejected over  0.25--1.5 AU and/or episodic magnetic tower
cavities launched from the disk appear as the most plausible origins
for the \feii\ medium velocity component in the DG Tau jet.
 The same disk wind model can also account for the properties of the high-velocity \feii\ flow, although alternative origins in magnetospheric and/or stellar winds cannot be excluded for this component.}
\maketitle
%

\section{Introduction}

The physical mechanism by which mass is ejected from accreting young
stars and is collimated into jets and the role that these jets play in
solving the angular momentum problem of star formation remain
fundamental open questions. It is now widely accepted that the action
of magnetic fields is required to explain the efficient jet
collimation and acceleration \citep{Cabrit2007_jetset}. However,
several MHD ejection sites may operate in parallel: the stellar
surface, reconnection points in the stellar magnetosphere, the
inner disk edge or a broader range of disk radii, and their relative
contribution to the jet remains subject to debate \citep[see e.g.][and references therein for a recent review]{Ferreira2006,Edwards2009}.

Distinguishing between these scenarii is crucial not only for
understanding the role of jets in star formation, but also for
planetary formation and disk evolution. For example, the equipartition
magnetic field required to launch a powerful MHD disk wind could
affect planetary migration \citep{Terquem2003} and would speed up the
accretion flow inside the launch zone, lowering the disk surface density by several orders of magnitude compared to a standard viscous
disk model \citep{Combet2008}.

Strong observational constraints on the jet driving mechanism in young
stars come from studies of extended fordidden line emission in
"microjets" from young low-mass ($\le 2 M_{\odot}$) T~Tauri stars
(TTS). Their proximity and lack of obscuring envelope give access to
inner jet regions within 20--200 AU where the flow structure has,
hopefully, not yet been significantly perturbed by interaction with
the ambient medium. In addition, the optically thin forbidden lines
provide powerful tracers of jet density and excitation. In particular,
a wealth of data on jet widths, kinematics, and density and ionization
structure has been obtained over the last 10 years in the optical
range, thanks to the subarcsecond resolution provided by adaptive
optics on the ground and HST/STIS \citep[see the review by][and references therein]{Ray2007}. Confronting proposed ejection models with these data,
\citet{Ferreira2006} proposed that centrifugal disk winds appear the
most promising to reproduce simultaneously all forbidden line jet
observations, and that these disk winds could confine the hot inner
stellar / magnetospheric winds inferred, e.g., from He~I line profiles
\citep{Kwan2007}.

 Support for an extended MHD disk wind is most evident in the
bright and well-studied jet HH~158 from the actively accreting T Tauri
star DG~Tauri \citep{Mundt1983}. An onion-like structure is
observed within 500~AU of the star in optical forbidden emission lines, with
faster gas nested inside slower material
\citep{Lavalley-Fouquet2000,Bacciotti2000}. This behavior is naturally
expected if the wind is launched from a broad range of disk radii.
Furthermore, the slowest jet material at $V_r \simeq 30-60$ \kms\
exhibits velocity shifts between the two sides of the jet
\citep{Bacciotti2000,Coffey2007} in the same sense as the rotation of
the DG Tau disk \citep{Testi02}. These transverse shifts 
excellently agree with predicted rotation signatures for a
steady-state MHD disk wind launched out to about 3 AU
\citep{Anderson2003,Pesenti2004,Ferreira2006}. A velocity shift in the
same sense is seen across faster jet material at -200 \kms\, suggesting
(if caused by rotation) ejection from smaller disk radii of $\simeq$
0.2--0.5 AU \citep{Coffey2007}.
 

However, rotation signatures alone are insufficient to firmly
establish a disk wind origin for the DG Tau jet, because velocity shifts
might be contaminated by other effects, such as jet precession or non-
axisymmetric shocks \citep{Cerqueira2006,Soker2005}. Alternative
explanations for the broader, slow component in forbidden lines could
be, e.g., a cocoon of gas ejected sideways from the central fast jet
beam or ambient swept-up gas. To further test these alternatives, more
information is needed such as the full flow morphology in 2D, accurate
mass-fluxes, and dust content. 

Another open question is whether the fast and slow material seen in
forbidden lines in the DG~Tauri jet share a similar origin or if they
trace two distinct ejection sites and ejection mechanisms. 
A dual
origin in a magnetospheric wind and a disk wind was proposed by
\citet{Pyo2003} for the emission in \feii from long-slit observations at
high spectral resolution (30 \kms) based on the double-peaked profile with a
high-velocity component at about -200 \kms\ and a
''low-velocity'' component  at about -100 \kms, of relative
strength varying along the jet. In contrast, a single disk wind where
the ionization fraction varies continuously from inner to outer
streamlines would predict a smooth single-peaked profile, as shown in
Fig.~ 2 of \citet{Pesenti2004}. However, the ionization distribution
in the disk wind has a crucial effect on the line profile shape, with,
e.g., the peak shifting from high to low velocity depending on whether
ionization decreases or increases outward
\citep{Pesenti2004}. Therefore, double-peaked profiles could in
principle be produced in a single disk wind if ionization conditions
are highly inhomogenous between inner and outer streamlines, e. g. because of
localized shocks. To better constrain the respective origins of the
two velocity components noted in \feii\ by \citet{Pyo2003}, additional
constraints are again needed.



In this paper, we present spectro-imaging of the DG Tau jet in the H-band
that provide the first images of this jet in \feii\ at $\simeq$
0\farcs15 resolution over a full field of view of $\pm$ 1\farcs5 from
the star. This enables us to clarify the morphology and opening angles
of the two \feii\ velocity components identified by \citet{Pyo2003} in
long-slit spectroscopy. We also identify new knots and bubble features
in both the jet and counterjet, and derive the electronic density
distribution along the jet, using the \feii\ 1.53\mic/1.64\mic\
ratio. The observations and data reduction are presented in
Section~\ref{sec:obs}, and our results in terms of morphology,
kinematics, collimation, red/blue asymmetry, disk occultation, and jet
densities are presented in Section~\ref{sec:results}. In
Section~\ref{sec:discussion} we use these results to revise the
timescale for knot formation in the DG Tau jet, measure iron depletion
in the two velocity components identified by \cite{Pyo2003} and
estimate their mass-flux and ejection to accretion ratio. These
results are  compared to proposed ejection processes in T
Tauri stars, in particular the X-wind model \citep{Shang1998} and the
extended MHD disk wind model invoked to fit the rotation signatures in
DG~Tau \citep{Pesenti2004}. Our conclusions are summarized in
Section~\ref{sec:conclusions}.


\section{Observations and data reduction}

\label{sec:obs}
\begin{figure}
\resizebox{\hsize}{!}{\includegraphics{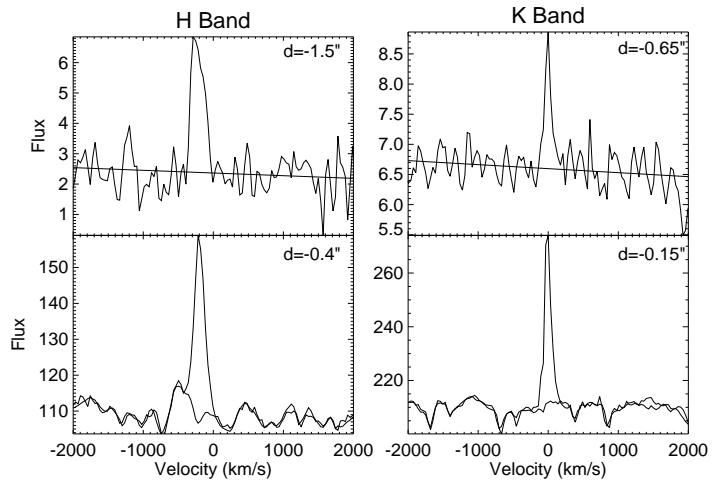}}
\caption{Illustration of the continuum substraction methods, close to the star (bottom row) and far from the star (top row), for the \feiil\ (left panels) and the \h2\ line (right panels). The raw spectra are plotted in thick lines and the fitted continua in thin lines. See text for details. The analysis and discussion of \h2\ lines will be presented in a separate paper.}
\label{fig:cont_sub}
\end{figure}

Spectro-imaging observations of the DG~Tau microjet in the H- and K-bands were conducted on 2005 October 15 at the Very Large Telescope
(VLT) with the integral field spectrograph SINFONI combined with an
adaptive optics (AO) module \citep{sinfoni1,sinfoni2}. We used a
spatial sampling of 0\farcs05~$\times$0\farcs1, with the smallest
spaxel dimension aligned  at PA = 315$^{\circ}$, i.e.  in the
direction transverse to the jet axis, denoted $x$-axis in the
following. This provided a total field of view of about
3\arcsec~$\times$ 3\arcsec. The spectral resolution was
$\Delta\lambda/\lambda \sim$3000 in the H-band and $\Delta\lambda/\lambda
\sim$4000 in the K-band, with a spectral sampling of 35 \kms /pixel and
 34 \kms /pixel in H and K  respectively. In the H-band, individual
short exposures of 15sec were coadded to build a final datacube with
total integration time of 2$\times$270s on the object and 270s on the
sky. Thanks to AO correction, the effective spatial resolution
achieved is $\sim$0.15\arcsec FWHM in the PSF core (estimated from the
reconstructed continuum image).

The data reduction process is common to the two spectral bands. Data
reduction steps were carried out with the SINFONI pipeline. Datacubes
were corrected for bad pixels, dark, flatfield, geometric distortions
on the detector, and sky background.  The combined datacube was resampled to
a 50 mas $\times$ 50 mas square spatial grid.  A raw wavelength calibration
based on daytime arclamp exposures is performed by the
pipeline. However, the accuracy of this calibration (on the order of
50\kms) is insufficient for our kinematical jet studies. For the
\feiil\ line, we improved this wavelength calibration locally using the
bright OH sky emission line at $\lambda$1.6692\mic\ to obtain a 2D
wavelength solution over our whole field of view. We estimate a final
calibration error in velocity of 10\kms~rms in this line. The
wavelength scale was converted into the stellar rest frame velocity
scale. Unfortunately, for the \feii $\lambda$1.53\mic\ line, nearby OH
lines have a signal-to-noise ratio that is too low to improve wavelength
calibration. Therefore, only velocity-integrated 1.53\mic\ line fluxes
are used in the analysis. The data were flux-calibrated on the A2V
standard star HIP~26686 in the H-band. We estimate an error in the
flux calibration of 5\%. Correction for telluric absorption lines was
performed together with our continuum subtraction routine (see below).
To correct for differential atmospheric diffraction and ensure proper
registering, the final datacube for each line of interest is spatially
recentered on the centroid position of the local continuum.

The bright stellar spectrum is strong over most of the field of
view. In particular, a strong photospheric line is located under the
\feiil\ transition. To better retrieve the intrinsic \feii\
jet emission profile, especially close to the central source, we
performed a specific continuum subtraction, illustrated in the bottom
row of Figure~\ref{fig:cont_sub}. We selected a high signal-to-noise reference photospheric spectrum free of jet emission in the SINFONI field of
view  (at $\Delta x$=0.054\arcs, $\Delta y$=-0.003\arcs).
This reference spectrum was then scaled down to fit the local continuum
level and was subtracted out from the current spectrum. The high quality
of this continuum-fitting procedure can be seen in the bottom row of
Figure~\ref{fig:cont_sub}. It allows us to retrieve the extended jet emission
features down to 0\farcs 2 from the star. Furthermore, the residual
emission line profile is not only free of photospheric absorption
lines but also of telluric features against the continuum, because the
depth of telluric absorption scales in proportion to the local flux
level\footnote{Our photospheric subtraction procedure would also
subtract any {\it unresolved} \feii\ feature centered {\em on} the
star. However, no such compact component appears to be present in
previous high-resolution \feii\ spectra \citep{Pyo2003}.}. Farther from
the star, where the line-to-continuum ratio becomes high, this
procedure only provides a partial correction for telluric
absorption. However, the telluric absorption depth is less than 7\%
locally around the \feii\ lines of interest, therefore the error in
the emission-line flux is comparable to the flux-calibration
uncertainties. At distant positions, where the signal to noise in the
stellar continuum is too low for proper continuum scaling, a simple
linear baseline is fitted to the local continuum around the line and
subtracted (top row in Figure~\ref{fig:cont_sub}).  At each
position, we also estimate a spectral noise from the standard
deviation of the residual (continuum-subtracted) spectrum on either
side of the line. The same continuum subtraction procedures were
applied to the H$_2$ 2.12\mic\ line in the K-band. The results in H$_2$
will be presented in a separate paper. 

\section{Results}
\label{sec:results}

\begin{figure*}[t]
\resizebox{\hsize}{!}{
\includegraphics[width=0.55\textwidth,trim= -0.2cm 0cm 7.5cm 0cm,clip=true]{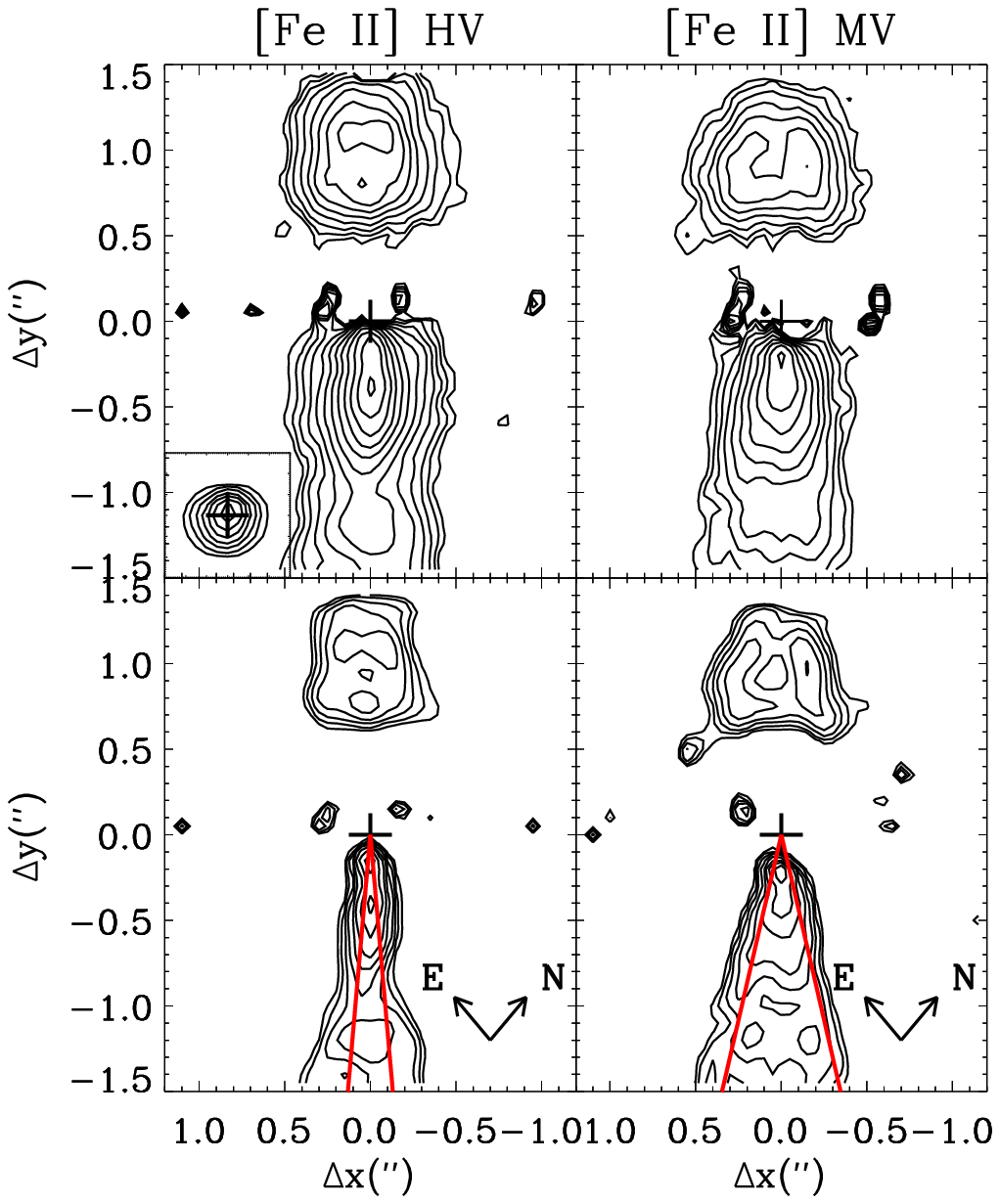}
\includegraphics[width=0.35\textwidth]{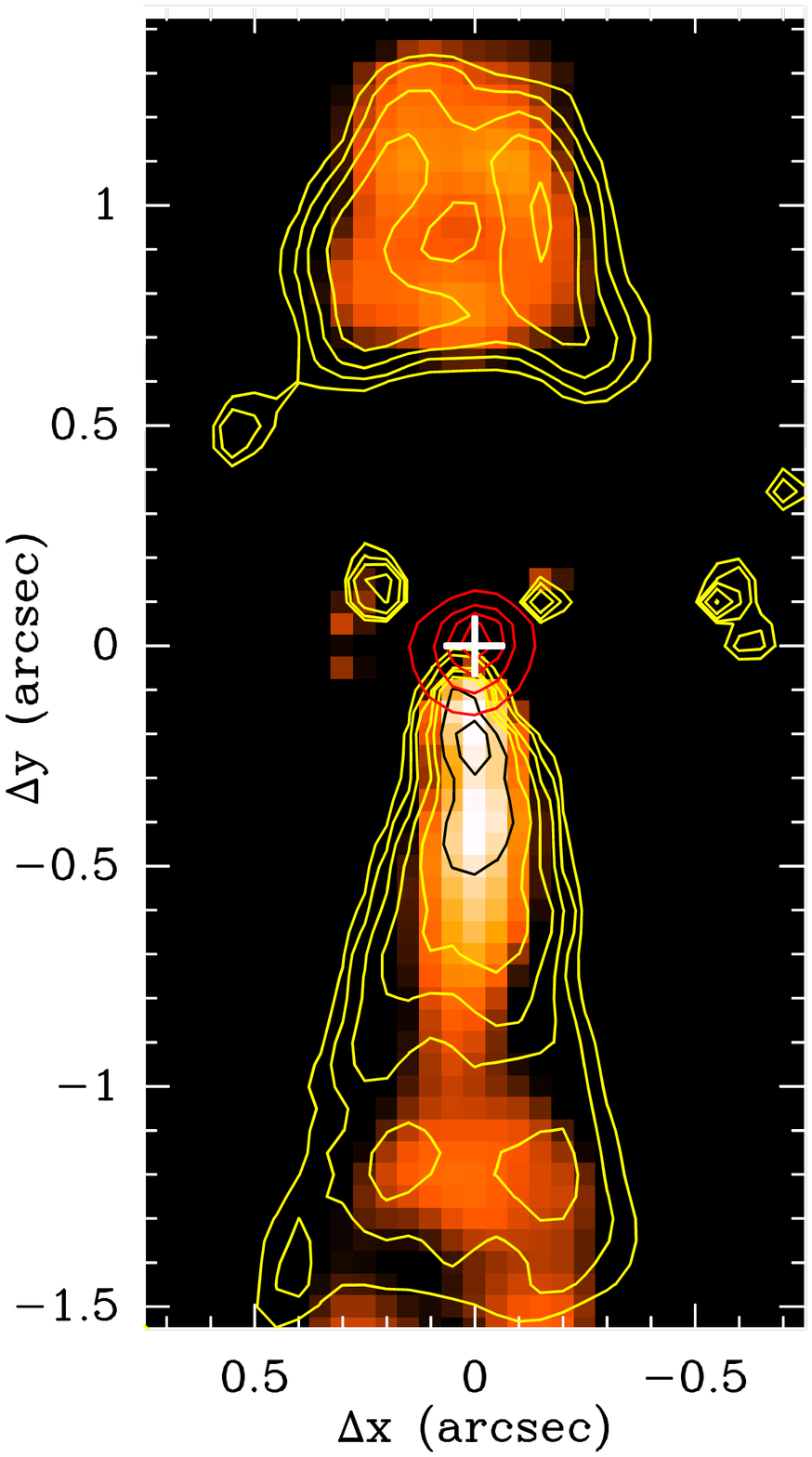}
}
\caption{ 
\sl Left \rm: Continuum-subtracted \feiil\ channel maps of the DG Tau
jet before (top panels) and after (bottom panels) deconvolution by the
continuum image. The HV channel maps are integrated
over the velocity intervals [-300,-160] \kms\ (HVB) and
[+120,+260] \kms\ (HVR); the MV channel map over [-160,-50] \kms\ (MVB)
and [-50,+120] \kms\ (MVR).  The blueshifted lobe is at $\Delta y <$~0. The pixel scale is 0\farcs05. 
A cross denotes the centroid of the continuum emission
(shown as an insert in the top left panel).  Contours  start at 3
$\sigma$ and  increase by factors of $\sqrt{2}$ (lowest contours
are 2.0 and 1.8 $\times 10^{-18}$ W m$^{-2}$ arcsec$^{-2}$
in the raw HV and MV maps respectively). The solid red
lines in the deconvolved maps show the opening angle measured in each
component.  \sl Right \rm: Superposition of the deconvolved HV
(background color image) and MV (yellow contours) channel maps. The
stellar continuum is shown in red contours. The white cross locates
the continuum centroid.
}
\label{fig:channelmaps}
\end{figure*}

The present H-band SINFONI data provide the first observations on sub-arcsecond scales of the 2D spatio-kinematic structure of \feii\ emission in the DG~Tau jet. Below, we present in turn several new results obtained from our data:
\begin{itemize}
\item the \feii\ jet morphology and collimation  in various velocity intervals
\item the kinematic structure and variability of \feii\ emission, compared to earlier observations in \feii\ and optical lines
\item the red/blue asymmetry and constraints on the occulting disk
\item the electronic density distribution derived from the \feii\ 1.53\mic/1.64\mic\ ratio.
\end{itemize}

\subsection{2D morphology in \feii} 

We show in the top row of Fig.~\ref{fig:channelmaps} the raw
continuum-subtracted \feii\ channel maps reconstructed from our
datacube. The emission in each jet lobe was separated in two
velocity ranges, chosen to correspond best to the two kinematical
components at -200~\kms\ and -100~\kms spectrally identified by
\citet{Pyo2003} in their 2001
\feii\ long-slit spectrum along the jet axis \footnote{We will use here the notation MV (medium-velocity) rather than LV for the component at $\simeq
-100$\kms\ in \citet{Pyo2003} to be more consistent with the velocity
intervals defined in the HST channel maps of \citet{Bacciotti2000}; we
restrict the use of LV to the component at velocities lower than
-50 \kms, seen in optical lines.}.  The velocity intervals are the
following: high-velocity blueshifted (HVB) = [-300,-160] \kms,
medium-velocity blueshifted (MVB) = [-160,-50] \kms, medium-velocity
redshifted (MVR) = [-50,120] \kms, high-velocity redshifted (HVR) =
[120,260] \kms. Because of our moderate velocity resolution and
the lower velocities in the redshifted jet, we extended the MVR channel
map to -50~\kms. The HV channel map shows the sum of the HVB and HVR
intervals. The MV channel map shows the sum of MVB and MVR. The
blueshifted lobe is at $\Delta$y $<$ 0  in Fig.~\ref{fig:channelmaps}.
 
The bottom row of Fig.~\ref{fig:channelmaps} shows the channel maps
after deconvolution by a continuum image reconstructed from the same
datacube, which provides a good estimate of the point-spread
function. We used the LUCY algorithm implemented in the
IRAF/STSDAS~V3.8 with 40 iterations. We checked that the derived jet
morphology (in particular the jet widths) did not change significantly
with further iterations. The right-hand panel of
Fig.~\ref{fig:channelmaps} presents a color superposition of the
deconvolved HV and MV channels to facilitate comparison
between them and to highlight faint emission features.

The deconvolved channel maps reveal several important new features: 

(1) the blueshifted \feii\ jet shows a clear increase in collimation
at higher flow velocities, with the HVB emission nested inside the MVB
emission. This onion-like velocity structure was previously noted in
optical lines \citep{Lavalley-Fouquet2000, Bacciotti2000}, but the
unique combination of high resolution and wide transverse coverage
provided by SINFONI shows it more clearly. In particular, it confirms
the striking conical shape of the MVB emission, which was partly
truncated in the narrower HST/STIS channel maps covering $\pm
0\farcs2$ from the jet axis \citep{Bacciotti2000}.

(2) The blueshifted jet contains several knots and bubble-like
structures: Two knots are present, one at  -0\farcs37 $\pm
0.03$\arcsec\ and one at -1\farcs2 $\pm 0.05$\arcsec. The latter is
followed by a faint limb-brightened bubble, partly truncated by our
field of view, which reaches a maximum width $\simeq$ 0\farcs5 around $
d \simeq$ -1\farcs4 (see right panel in Fig.~\ref{fig:channelmaps}).

(3) The counterjet emission is dominated by a striking limb-brightened
circular bubble centered around $\simeq +1$\arcsec. The deconvolved
MVR emission is slightly broader and more pointed away from the star
than the HVR.
No counterjet is detected within 0\farcs7 from the star. This effect
is probably caused by disk occultation \citep{Lavalley1997,Pyo2003}
and provides interesting constraints on the disk surface density
structure in DG Tau (see Section~\ref{sec:counterjet}).

\subsection{Collimation, opening angle, and upper limits on launch radii}

\begin{figure*}[t] 
\includegraphics[angle=270,width=0.48\textwidth]{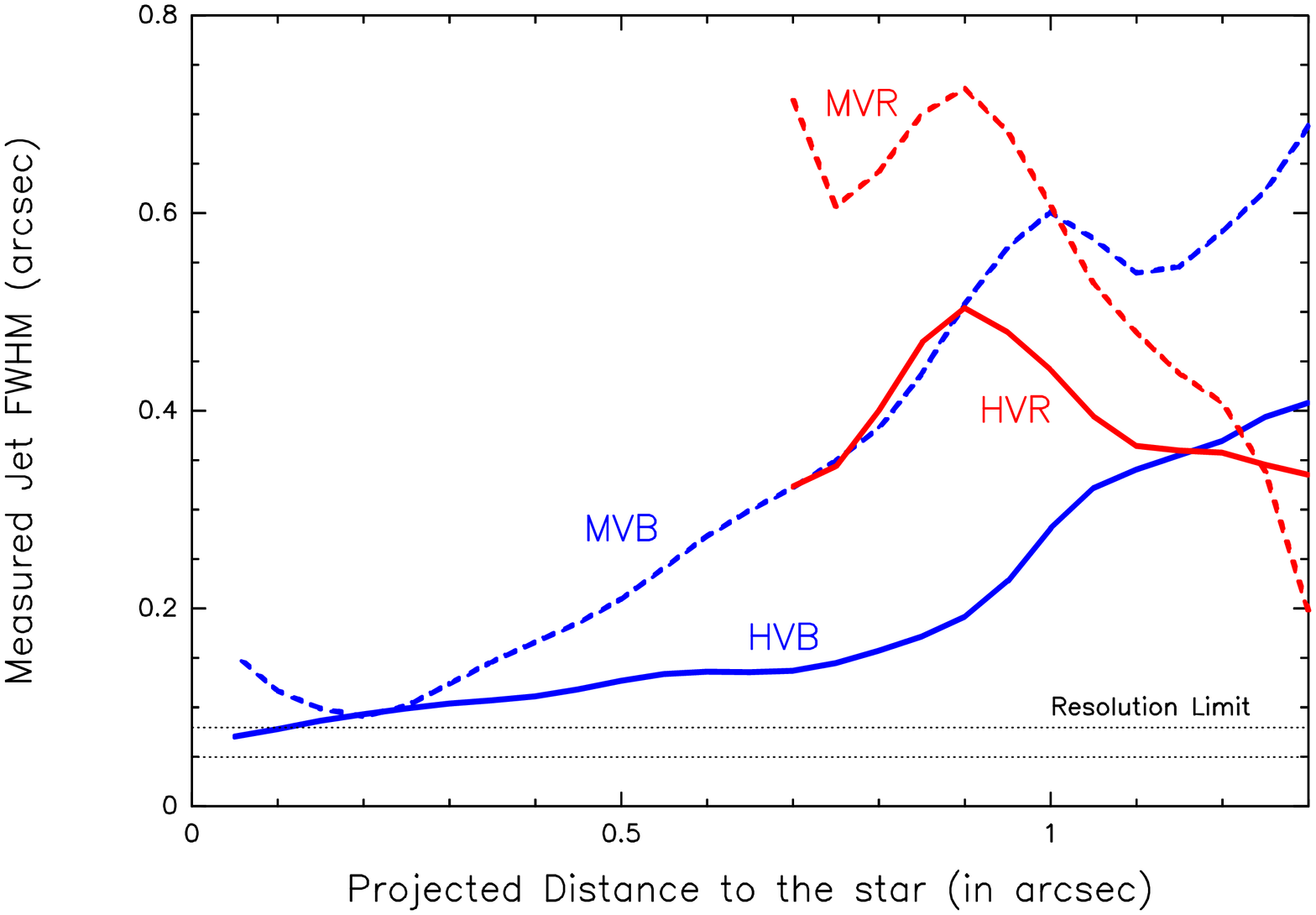}
\includegraphics[angle=270,width=0.48\textwidth]{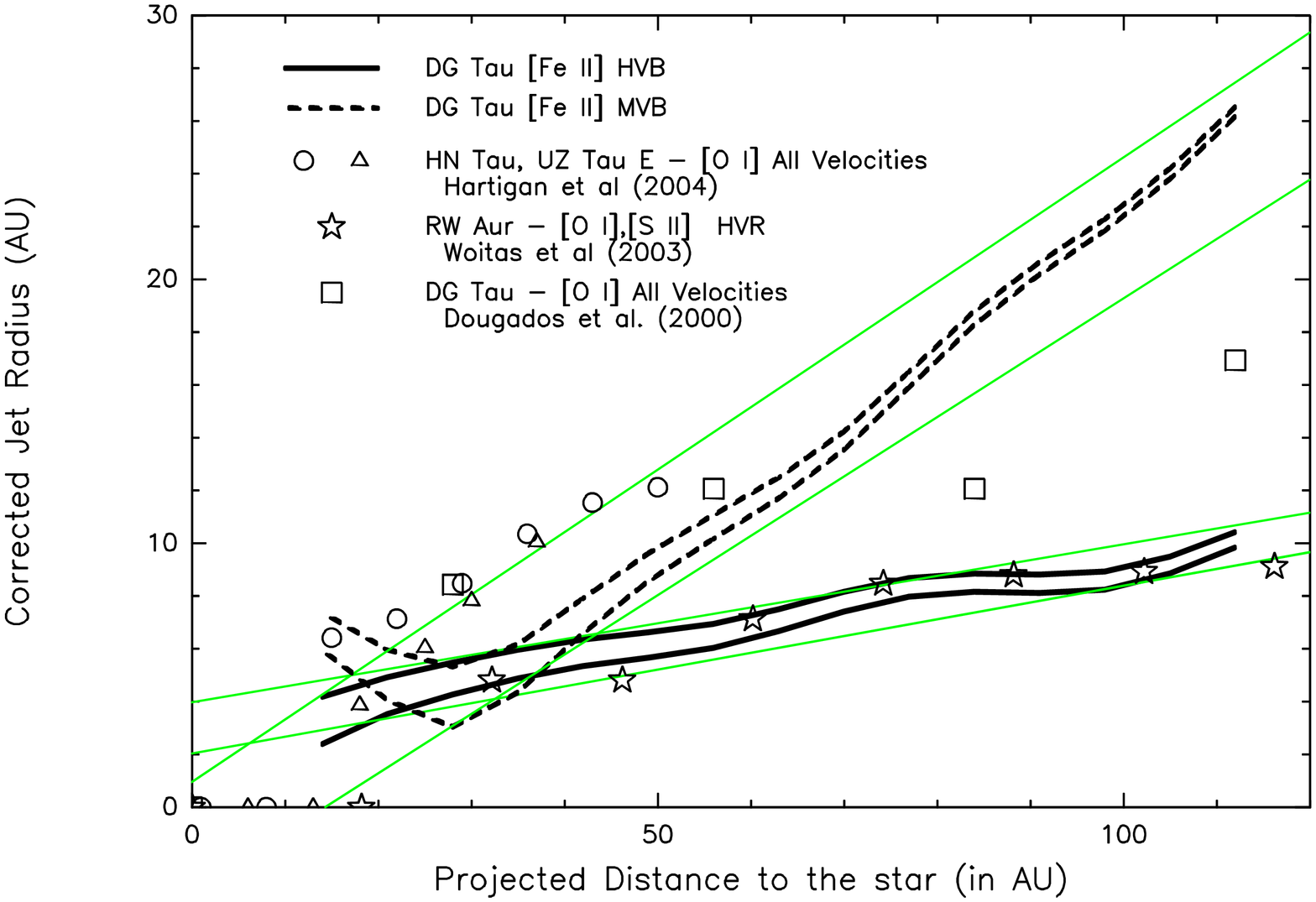}
\caption{{\sl Left}: Jet FWHM derived from Gaussian fits to deconvolved
\feii\ channel maps as a function of projected angular distance from
the star. {\sl Right}: PSF corrected (i.e. intrinsic) jet radii as a function 
of linear
distance from the star for the inner 0\farcs8 of the HVB (solid
curves) and MVB (dashed curves). The fitted FWHM across the jet were
 corrected for the angular resolution of 0\farcs05--0\farcs08 in
the deconvolved maps by subtracting this PSF in
quadrature. Gray/green solid lines show fits by a flow of constant
opening angle, suggesting a launch radius smaller than 4~AU (HVB) and
1~AU (MVB). For comparison, open symbols plot jet radii measured in
\oi\ images integrated over all velocities in DG Tau \citep[open squares][]{Dougados00}, RW Aur \citep[open stars][]{Woitas2002}, and HN
and UZ Tau E \citep[open circles and triangles][]{Hartigan2004a}.}
\label{fig:jetwidths}
\end{figure*}

We plot in the left panel of Fig.~\ref{fig:jetwidths} the jet FWHM derived from Gaussian fitting of the transverse intensity profiles in the deconvolved channel maps of Fig.~\ref{fig:channelmaps}. The broader width of the MV channels compared to the HV is clearly visible at all distances. In addition, a widening of the blue jet is seen beyond 1\arcsec, corresponding to the faint bubble feature at the end of this lobe. The width there is comparable to the bubble feature in the redshifted lobe.  

To estimate intrinsic jet widths and opening angles, we focused on the
inner 0\farcs8 (110 AU) section of the blueshifted lobe, which is
least affected by bubble structures. We corrected the measured FWHM for
the effective PSF width ($FWHM_0$) in the deconvolved channel maps
with the formula (2R$_{\rm jet})^2 = {FWHM^2-FWHM_0^2}$. The effective
angular resolution reached in the deconvolved maps cannot be measured
precisely because no unresolved point source is present in our
(continuum-subtracted) line maps. Therefore we considered a range of
possible $FWHM_0$ between 0\farcs05 (pixel scale) and 0\farcs08
(smallest jet width measured). We plot in the right panel of
Fig.~\ref{fig:jetwidths} the PSF-corrected jet radii (in AU) in the
HVB and MVB channels as a function of distance as well as linear fits
to these data points.

We find that the medium-velocity blueshifted (MVB) emission has a
conical shape with a roughly constant half opening angle of
14$^{\circ}$. The high-velocity blueshifted emission (HVB) keeps a
narrow radius (5--10 AU) with a half opening angle of 4$^{\circ}$ until
0.8\arcs\ from the star. The derived opening angles are upper limits
because we did not correct for inclination effects. {\it If these cones
traced actual streamlines}, an extrapolation back to the origin would
suggest that the HVB component in DG Tau originates in a disk radius $<$
4~AU and the MVB in a disk radius $<$ 1~AU. This is compatible with the
maximum launch radii of 0.5~AU and 2--3~AU derived from
tentative jet rotation signatures at projected flow speeds of -200
\kms\ and $\simeq$ -50 \kms\ respectively \citep{Coffey2007, Bacciotti2002,
Pesenti2004}. However, it is likely that emission maps in constant velocity bins do not trace actual streamlines especially at low flow velocities.
Indeed, outer streamlines are expected to reach their terminal velocities on spatial scales $\simeq$ 50-100~AU. As a consequence, jet emission radii measured on the MVB channel maps likely underestimate
true streamline radii close to the source.

We also compare in Fig.~\ref{fig:jetwidths} our derived intrinsic jet
radii in \feii\ with previous measurements from {\em
velocity-integrated} \oi\ images at similar angular resolution in the
RW Aur jet (average of both lobes) \citep{Woitas2002} and in the HN
Tau and UZ~Tau~E blueshifted jets \citep{Hartigan2004a}. The RW Aur
microjet, where \oi\ is dominated by a single high-velocity component
\citep{Woitas2002}, shows an opening angle and width strikingly
similar to the HVB interval in DG Tau. The HN Tau and UZ Tau E \oi\
jets, dominated by a medium to low-velocity component
\citep{Hartigan1995}, have a much wider opening angle similar to the
MVB. Therefore, the increase in collimation at higher velocity seen in
DG Tau could be a widespread generic property in T Tauri
jets. Finally, we note that the jet widths from a velocity-integrated
\oi\ image of the DG Tau jet taken in 1997 \citep{Dougados00} fall
close to the MVB component out to 50 AU, but closer to the HVB beyond
50 AU. Therefore, jet opening angles measured from velocity-integrated
images may be misleading when different velocity components dominate
at different positions along the jet.

\subsection{2D kinematics of the blueshifted jet}
\label{sec:kinematics} 

\begin{figure*}
\includegraphics[width=0.45\textwidth]{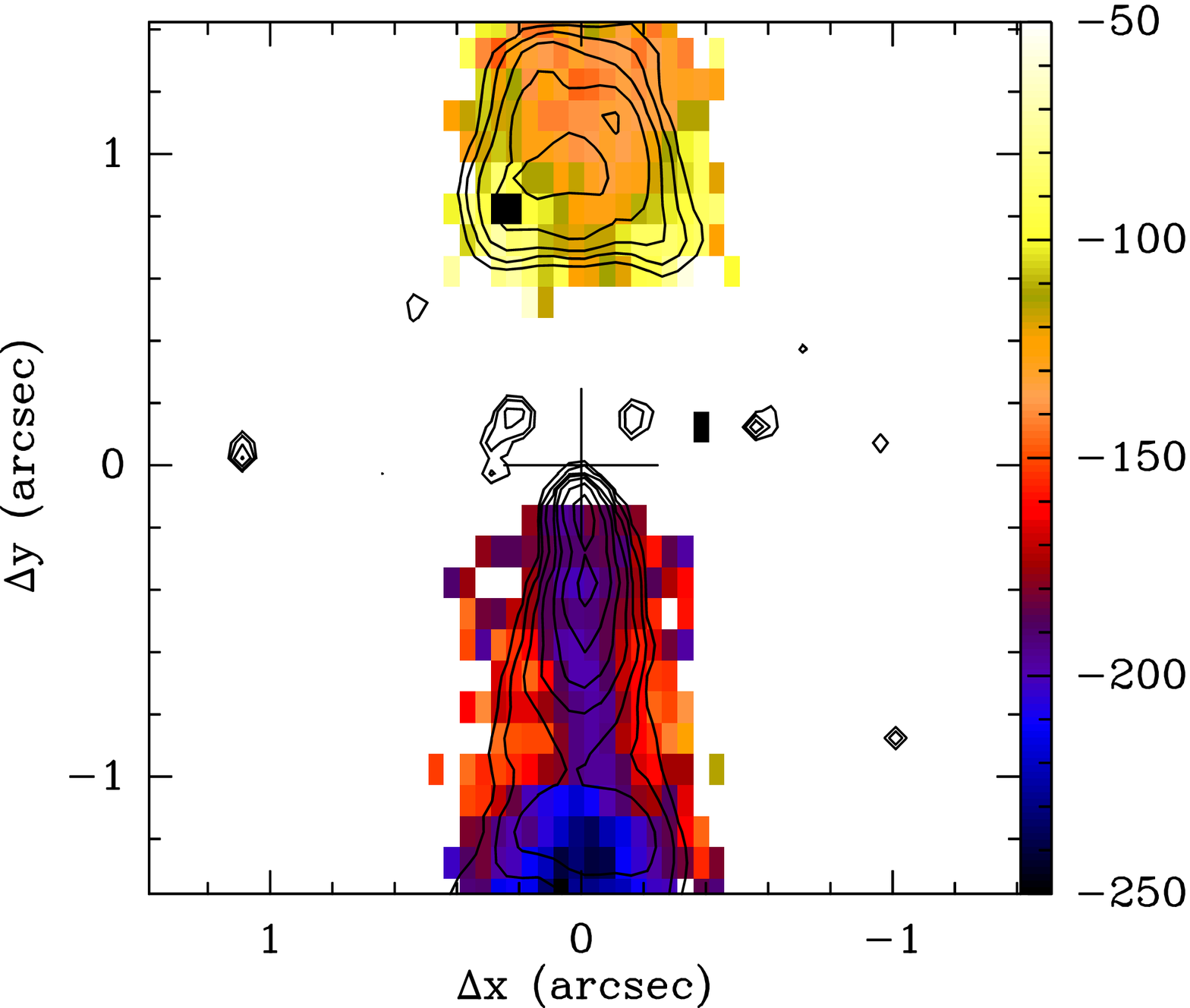} 
\includegraphics[width=0.55\textwidth]{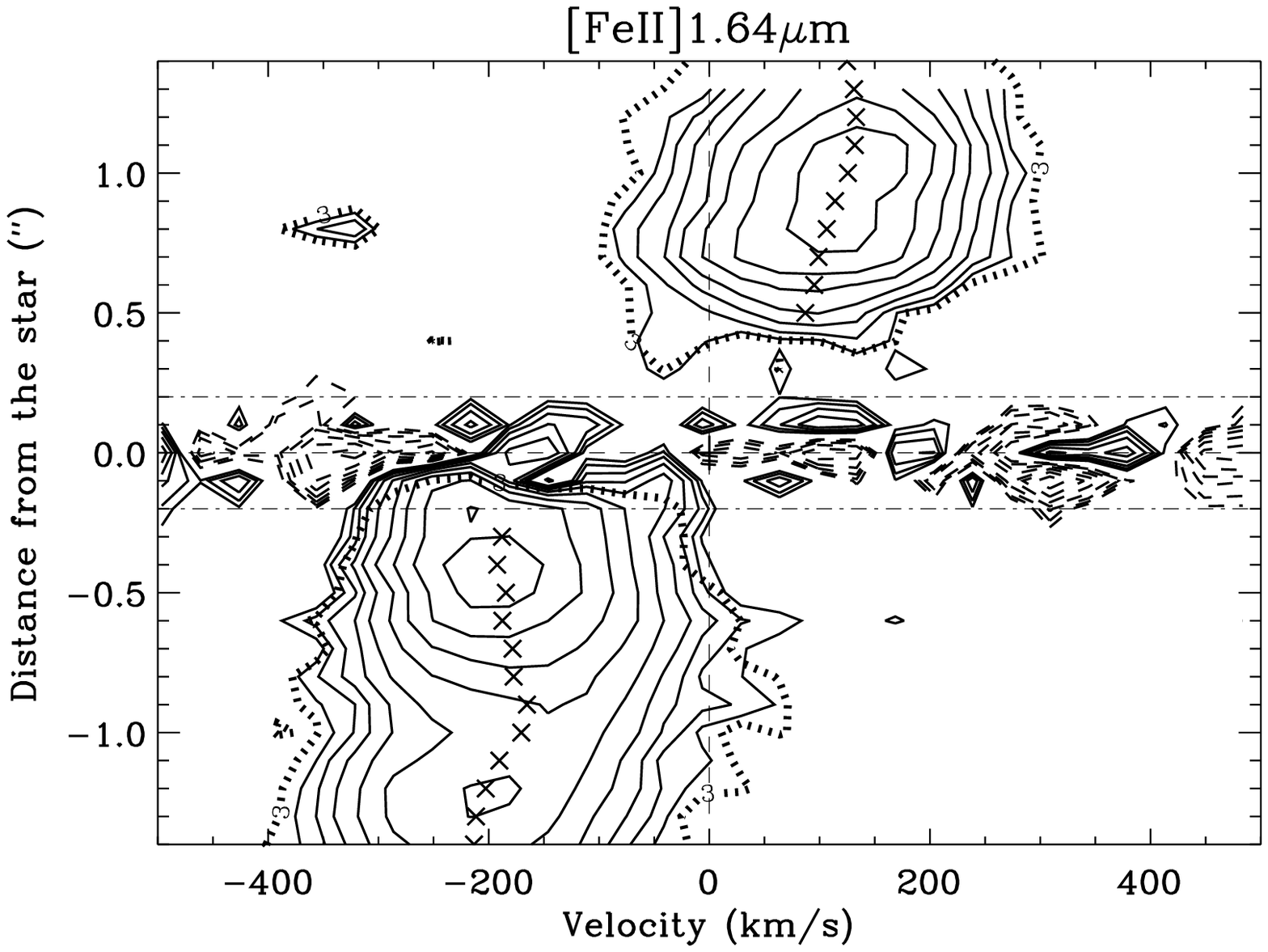} 
\caption{Left: 2D maps of the \feiil\ velocity centroid deduced from one component Gaussian fitting of the line profile.  Redshifted velocities are shown on a negative scale to better highlight the velocity asymmetry between them. Spectra with SNR lower than 5 were masked. The cross marks the location of the continuum.  Contours show the sum of the deconvolved HV and MV channel maps.  Right: PV diagram of \feii\ along the jet, averaged over $\pm 0\farcs5$ across the jet, with velocity centroids indicated as crosses. Horizontal dash-dotted lines indicate the region where residual noise after continuum subtraction is still large. Contours start at 2.43$\times$10$^{-15}$ W m$^{-2}$ $\mu$m$^{-1}$ arcsec$^{-2}$ and increase by factors of $\sqrt{2}$. The 3$\sigma$ contour is shown as a dotted line.}
\label{fig:velmapFeII}
\end{figure*}

To attain an overall view of the 2D jet kinematics, we performed at each spatial position a one-component Gaussian fit to the line profile. Although two kinematical components are clearly present in the jet (as shown by the distinct morphologies of channel maps in the HV and MV velocity ranges), a two-component Gaussian fit did not give coherent results over the field of view because of the moderate spectral resolution of our observations. We show in Fig.~\ref{fig:velmapFeII} the derived 2D map of the fitted centroid velocities. From the formula outlined in \cite{Agra-Amboage2009} and the observed signal-to-noise ratios, we estimate an uncertainty in centroid velocities in the range 14-21 \kms.

The centroid map shows a clear transverse velocity gradient in both lobes, with the highest flow velocities concentrated toward the axis. The gradient is especially apparent in the blue lobe at distances greater than 0.4\arcsec\ from the star. This transverse gradient is consistent with the higher degree of collimation in the HV channel maps and the wider, limb-brightened morphology in the MV channel maps. The centroid values toward the flow edges indicates a mean absolute velocity of $\le 50$ \kms\  for the MVR component, and
$\le 100-150$ \kms\ for the MVB component toward the blue  lobe, similar to the low velocity spectral component identified by \cite{Pyo2003} in their 2001 long-slit \feii\ spectrum, of higher spectral resolution than our data.  

Variations of centroid velocities along the jet axis are also
seen. They are illustrated in Fig.~\ref{fig:velmapFeII}, right panel,
where we plot a position-velocity diagram along the jet, reconstructed
by averaging our spectra over a 1\arcsec\ wide
pseudo-slit. Transverse-average centroid velocities decrease from -192
\kms at -0.4\arcsec\ (position of the first knot) down to -150\kms at
-0.8\arcsec\ and increase again to reach -210 \kms\ at the outer knot
at -1.2\arcsec.  We note a striking similarity with the amplitude and
period of the velocity variations in the high-velocity range reported
by \cite{Pyo2003} along the jet, despite the time difference of four
years between our observations.  When corrected for a jet inclination
of 42\degr, the variation amplitude $\Delta V_{\rm rad} \simeq$ 40-60
\kms\ suggests internal shock speeds $v_s \simeq 60-80$\kms, similar
to the $v_s \simeq 50-70$\kms\ inferred in 1998 from optical line
ratios in the velocity range [-250,-100] \kms
\citep{Lavalley-Fouquet2000}. In both lobes, the maximum radial
velocity is found toward the bubble feature near the edge of our
field of view.

One notable difference between our \feii\ PV diagram and that of \cite{Pyo2003}
 is that the line flux is dominated by the HVB all the way down
to -0\farcs2 from the star in our observations
(Fig.~\ref{fig:velmapFeII}, right panel), whereas it was dominated by
the MVB within 0\farcs5 of the star in 2001 \citep{Pyo2003}.  This
difference could be caused by the bright high-velocity knot present at
-0\farcs4 in our SINFONI map, which had not yet --- or barely ---
emerged in 2001 October (see Section~\ref{sec:knots}). Another difference is the
nondetection of the MVB component beyond 0\farcs5 from the star by
\citet{Pyo2003},which  may be caused by the narrow slit (0\farcs3 width) used in
their observations. As shown in Figure~\ref{fig:jetwidths}, the
measured MVB FHWM exceeds 0\farcs3 beyond 0\farcs6, therefore a
significant fraction of the MVB flux has likely been missed at these
distances in the \citet{Pyo2003} long-slit observations.
 
Finally, we note that neither our PV diagram nor that of Pyo et
al. from 2001  show \feii\ emission at velocities faster than
-300~\kms or lower than -60 \kms, corresponding respectively to the
"very high velocity" (VHV) and "low velocity" (LV) components
detected in the DG Tau blueshifted jet in optical lines respectively
\citep{Lavalley1997,Lavalley-Fouquet2000, Bacciotti2000}. 
HST maps obtained in 1999 show that the VHV optical gas was
tracing a bright knot followed by an elongated bubble
\citep{Bacciotti2000}. It was therefore produced by an intrinsically
transient event, and it is conceivable that new knots probed by the
2001--2005 \feii\ observations may not have reached these high
velocities.

\begin{figure}
\includegraphics[width=0.5\textwidth]{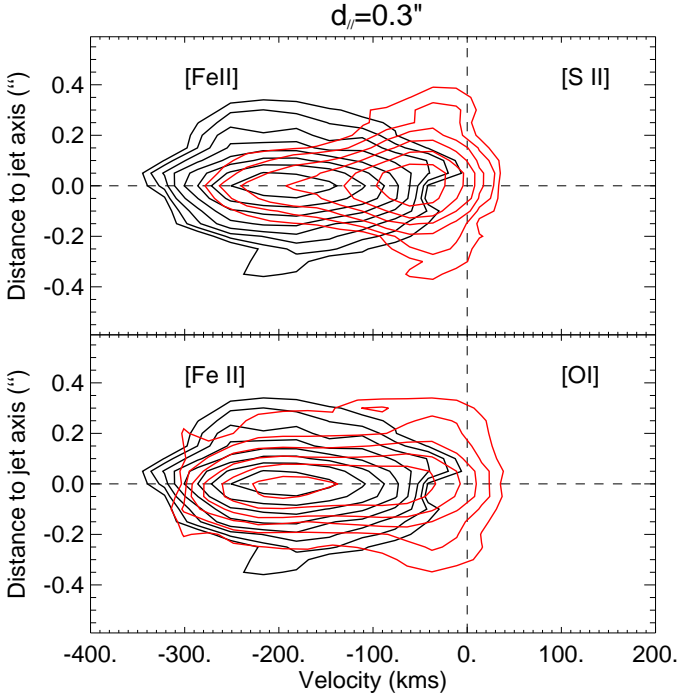} 
\caption{Comparison of transverse PV diagrams of \sii 6731\AA~ and \oi 6300\AA~ obtained in 2003 with HST/STIS at d=-0\farcs3 from the source in a 0\farcs1 wide slit \citep{Coffey2007,Coffey2008} with a transverse PV diagram in \feii\ at the exact same position, reconstructed from our SINFONI datacube obtained in Oct. 2005. Note the striking lack of \feii\ emission in the spatially wide low-velocity \oi\ component at V=-30,-40 \kms.}
\label{fig:pvtrans}
\end{figure}

On the other hand, the halo of broad LV emission seen in \oi\ and
\sii\ surrounding the blueshifted DG Tau jet is present in all optical
observations since at least 1993
\citep{Lavalley1997,Lavalley-Fouquet2000,Bacciotti2000} and most
recently in 2003 \citep{Coffey2007}.  Therefore it appears to be an
intrinsically stable feature, and its absence in \feii\ is
significant. To illustrate this, we compare in
Figure~\ref{fig:pvtrans} the transverse PV diagrams of \sii\ 6731\AA~ and
\oi\ 6300\AA~ obtained on 2003 December 1 with HST/STIS at -0\farcs3 from
the source \citep{Coffey2007} with a transverse PV diagram in \feii\
covering the same area, reconstructed from our SINFONI datacube. In
optical data, a spatially broad LV component is clearly visible of the
centroid decreasing from -60 \kms\ on-axis to -30,-40 \kms\ at $\pm
0\farcs2$ from the jet axis \citep{Coffey2007}. Similar low speeds
were observed at least out to 0\farcs6 from the star 
\citep[see parallel slits S1 and S7 in Fig. 3 of][]{Bacciotti2002}. The
weakness/absence of \feii\ emission from this "off-axis" LV component
is striking.  In Section~\ref{sec:discussion} we interpret this
difference as substantial iron depletion onto dust grains in the 
low-velocity sheath surrounding the jet.
 
\subsection{The counterflow: velocity asymmetry and constraint on disk structure}
\label{sec:counterjet}

The counterjet emission in our data is dominated by a striking
bubble-like structure of similar width and brightness as the bubble
partially imaged in the blueshifted jet (see
Figs.~~\ref{fig:channelmaps},\ref{fig:jetwidths}). On the other hand,
the velocity centroid map presented in Fig.~\ref{fig:velmapFeII}
reveals a clear velocity asymmetry between the red and blue bubbles.
The maximum measured radial velocity centroid toward the red bubble is
$\simeq$ +160 \kms, ie. only 70\% of the peak velocity $\simeq$ -230
\kms\ in the blue bubble. This velocity asymmetry is consistent with
the 70\% smaller distance from the star to the center of the
redshifted bubble, compared to the (approximate) center of its
blueshifted counterpart ($\simeq$ +1\arcsec\ versus -1\farcs4).  This
agreement suggests that both bubbles were generated in a simultaneous
ejection event.
 
The red/blue asymmetry in velocity and position needs to be taken into
account when quantifying the dust extinction produced by the DG Tau
disk toward the counterjet.  In October 2001, \citet{Pyo2003} detected
only very weak\footnote{the redshifted bubble was probably at that
time entirely hidden behind the high-extinction region close to the
star. Assuming a proper motion of 70\% that of the blueshifted knots
($\simeq$ 0\farcs3/yr, see Section~\ref{sec:knots}) the tip of the
redshifted bubble projected at 1\farcs3 arcsec from the star in
October 2005 would have been at 0\farcs46 in October 2001.} redshifted
\feii\ emission around 0\farcs9, 10 times weaker than their
blueshifted jet at the same distance from the star. They inferred an
extinction of 2.5 mag at H at this distance, assuming intrinsically
equal brightnesses.  If one corrects for the differential speeds and
compares instead with their blueshifted jet intensity at 0\farcs9/0.7
= 1\farcs3, the relative intensities become comparable, suggesting
negligible extinction by the disk at 1\arcsec.  This result is much
more consistent with the similar brightness of the red and blue
bubbles in our SINFONI data.

The deconvolved \feii\ channel maps suggest that the central
obscuration of the redshifted jet from the disk extends to projected
angular distances of d $\simeq$ 0.7$^{\prime\prime}$, i.e. 140~AU in
deprojected disk radius (assuming a distance of 140~pc and a disc
inclination of 42\degr).  This requirement appears to set constraints
on models of the disc density distribution in DG~Tau derived by
\cite{Isella2010} from sub-arcsecond millimeter continuum
interferometric maps. In particular, it seems inconsistent with a
single power-law surface density distribution, which requires a small
outer disk radius of 70-80~AU to fit DG Tau observations
\citep{Isella2010}. Such a small disk radius would produce no
extinction beyond 0\farcs4 from the star, unlike SINFONI
observations. An alternative similarity solution with an exponential
tail fitted to the mm map predicts a (perpendicular) surface density
of $\Sigma$=0.25-1.5~g~cm$^{-2}$ at a disk radius of r=100~AU
\citep{Isella2010}, leading to
$A_H = 30-5$ mag at d=0.5$^{\prime\prime}$ assuming interstellar dust
extinction, with $\Sigma$ and therefore $A_H$ dropping by a factor
$\simeq$ 10 at d=0.7$^{\prime\prime}$ (r=140~AU). The similarity
solution for the surface density distribution therefore appears more
consistent with the central obscuration observed in the SINFONI maps.

\subsection{Electron density in the \feii\ emitting region}
\label{sec:ne}

\begin{figure}
\centering
\includegraphics[width=0.5\textwidth]{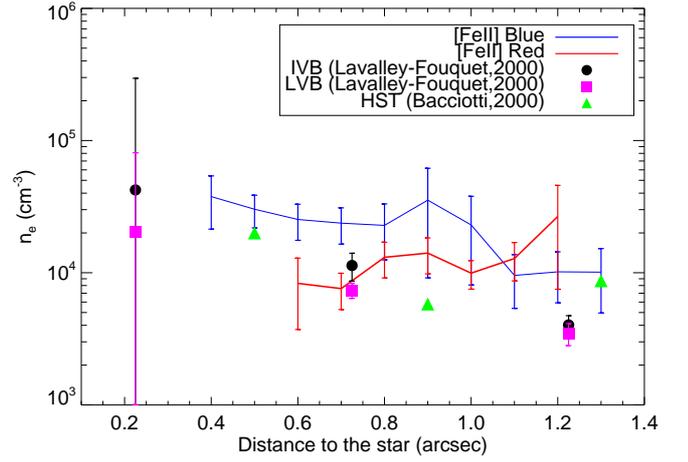} 
\caption{Plot of \ne\ versus distance derived from the ratio of the
\feii\ 1.53\mic\ to 1.64\mic\ lines (averaged over $\pm 0\farcs5$ from
the jet axis) for the blueshifted (blue points, top curve) and redshifted (red
points, bottom curve) lobe. Also shown are previous determinations from the \sii\
doublet ratio over similar velocity ranges: IVB and LVB from
\citet{Lavalley-Fouquet2000} (black circle, purple square), MVB interval of
\citet{Bacciotti2000} (green triangles).}
\label{fig:ne}
\end{figure}

The ratio $R$ of the \feii\ 1.53\mic\ to 1.64\mic\ lines is a good
diagnostic of the electronic density \ne\ in the range
$10^3-10^5$~cm$^{-3}$ \citep{Pradhan1993}. The temperature dependence
of $R$ is weak in the range $T_{ \rm e}\simeq$ 8000 - 20,000~K
encountered in the forbidden-line emission regions of stellar jets
\citep[e.g.][]{Bacciotti1999}. We did not detect any change in this ratio
across the line profile nor across the jet width at our
resolution. Therefore we computed at each spatial distance along the
jet a mean \feii\ line ratio integrated over all velocities and over
$\pm$0\farcs5 from the jet axis.  The flux for each line was obtained by
fitting it to a Gaussian and then integrating over all
velocities. Errors in the Gaussian parameters are derived taking into
account observational errors on the spectral profile and propagated to
obtain the final errors on the derived ratio. We then derived the
electronic density from the observed line ratio using the theoretical
relation between $R$ and \ne\ at a temperature \Te\ $\simeq$ 10,000~K,
calculated from the 16-level Fe$^+$ model of \cite{Pesenti2003}. For
ratios $R$ between 1.0 and 3.9, this relation can be fitted
analytically within an accuracy of 20\% on \ne\ by

\begin{equation}\label{eq:ne}
n_{\rm e}(R) = 1.2 \times 10^4 \times (R - 0.035)\,/\, (0.395-R) \quad {\rm cm}^{-3}.
\end{equation}

The \ne\ values as a function of distance along the jet are presented
in Fig.~\ref{fig:ne}. For comparison, we also plot in
Fig.~\ref{fig:ne} earlier \ne\ determinations obtained in
January 1998 and 1999 from the optical \sii\ 6716,6731\AA\AA\ doublet ratio
over a similar velocity range: namely, the IVB [-250,-100] \kms\ and
LVB [-100,+10] \kms\ intervals of \cite{Lavalley-Fouquet2000}, and the
MVB [-195,-72] \kms\ interval of \cite{Bacciotti2000}. The HVB
[-319,-195]\kms\ interval of Bacciotti et al. had a saturated ratio
indicating electronic densities $\gg 10^4$\cm\ out to 1\farcs4.

This comparison shows that the overall \ne\ distribution along the
blueshifted DG Tau jet has remained remarkably similar in the velocity
range [-300,+10] \kms\ between 1998 and 2005, with a general increase
of only a factor 2 over this period.  The actual change in total flow
density, $n_H$, would be even smaller than a factor 2 if the
ionization fraction had increased over this time period or if the
\sii\ 6716,6731\AA\AA\ doublet were biased toward less dense jet
regions than \feii, as found by \citet{Nisini2005}. We will take
advantage of this small variability level to estimate iron gas-phase
depletion in Section~\ref{sec:depletion}.

\section{Discussion}
\label{sec:discussion}
Below, we analyze our SINFONI results and discuss the resulting implications concerning

\begin{itemize}
\item the timescale for knot and bubble formation 
\item the gas-phase iron depletion in the blueshifted jet
\item the jet mass-flux and ejection to accretion ratio 
\item the proposed ejection processes and launch radii in DG Tau.
\end{itemize}

\subsection{Period of knot generation in the DG Tau jet}
\label{sec:knots}
\begin{figure}
\centering
\includegraphics[height=0.5\textwidth,angle=270]{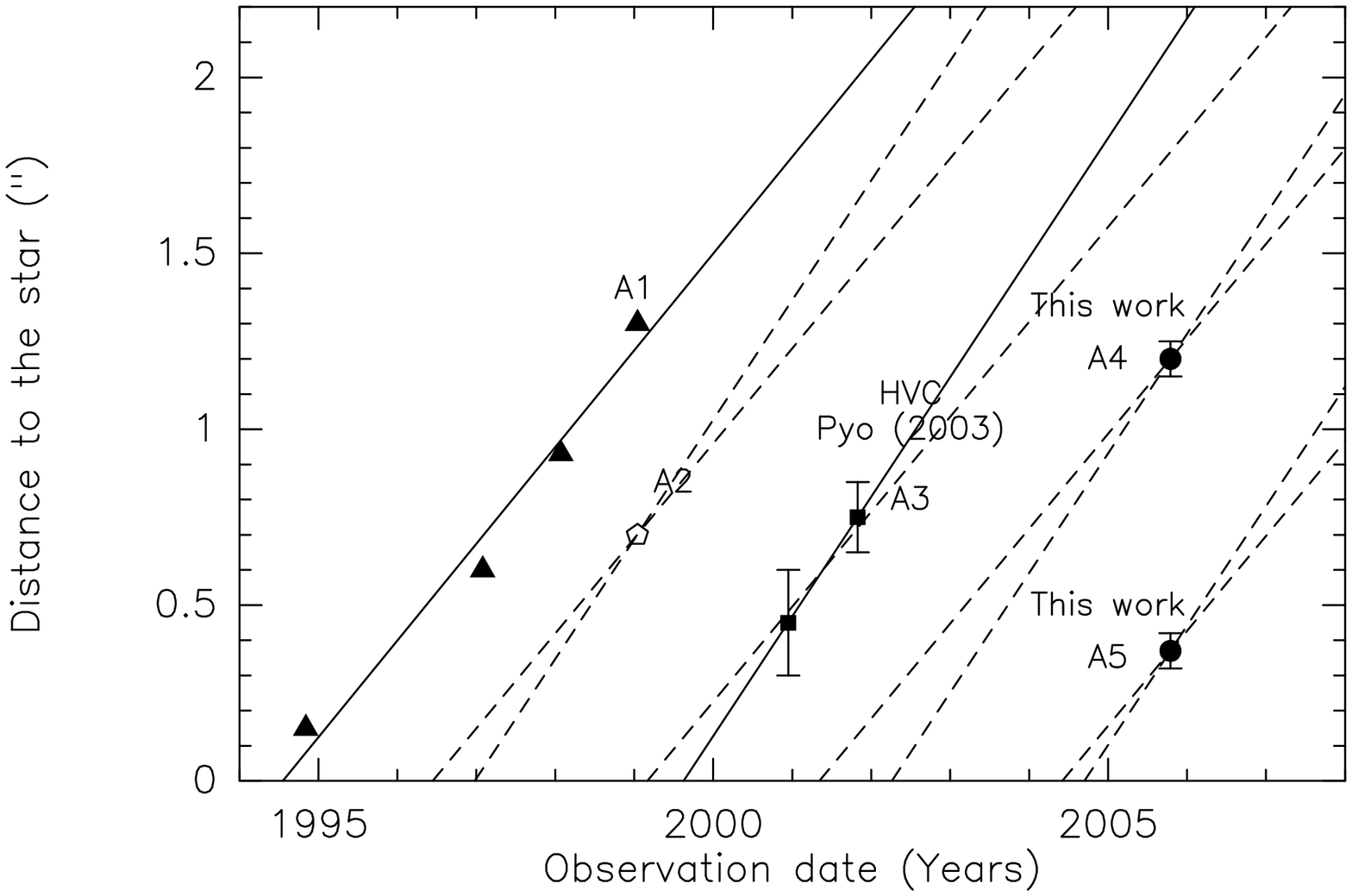}  
\caption{High-velocity knot position versus observing date in the DG Tau blueshifted jet updated from Fig.~5 of \citet{Pyo2003}. Solid lines show the proper motions derived by \citet{Pyo2003} for knots A1 and A3. For knots with only one epoch, dashed lines show the typical proper motion range of 0\farcs27~yr$^{-1}$--0\farcs34~yr$^{-1}$. The two new knots observed with SINFONI and knot A2 observed with HST indicate that a new knot typically emerges every 2.5 years since 1991. References for knot positions may be found in Table~1 of \citet{Pyo2003}.}
\label{fig:knots}
\end{figure}

 \citet{Pyo2003} investigated the knot proper motions and ejection timescales in the DG~Tau blueshifted lobe from 1991 to 2001, using a compilation of data from the literature. They concluded that a new knot emerges every five years.  Here we re-examine this matter and show that the SINFONI data indicate half that timescale.

The HVB emission map reveals two well defined \feii\ emission knots 
at distances from the star of -0\farcs37 $\pm$ 0.03\arcsec and  $-1\farcs2 \pm
0\farcs05$. In Figure~\ref{fig:knots} we add these
two HVB knots onto the distance-date diagram of \citet{Pyo2003} (their
Fig.~5). We also plot the proper motions determined by \citet{Pyo2003}
for knot A1 (0\farcs275 yr$^{-1}$ from four epochs) and for their HV peak
(0\farcs34~yr$^{-1}$ from two epochs). We denote the latter A3,
extending the knot-naming convention introduced by
\citet{Bacciotti2000}.

Evidently, the two high-velocity knots
identified in our HVB map correspond to new ejection events: the HV
knot A3 should lie at $\simeq$ -2\arcsec\ at the time of our
observations, i.e. well outside of the SINFONI field of view. If we
assume for these new knots A4 and A5, and for the previous knot A2,
proper motions in the same range 0\farcs27-0\farcs34~yr$^{-1}$ as
found for A1 and A3, we derive intervals between knots of about 2.5yrs
between 1995 and 2005 (typical spacing between knots of 0.75 \arcsec\
in the blue lobe). Our observations therefore suggest the presence of
a knot variability timescale  a factor two shorter than derived by
\citet{Pyo2003}.  We note that the bubble structures observed in our
SINFONI images appear to have formed in the wake of the same ejection
event as HV knot A3: if we adopt a proper motion in the red lobe of
70\% of the mean proper motion of 0\farcs3~yr$^{-1}$ in the blue lobe,
as suggested by the red/blue asymmetry (see
Section~\ref{sec:counterjet}), the position of the tip of the red
bubble at +1\farcs3 from the star in October 2005 yields an ejection
date around August 1999, the same as for knot A3. The formation of a
bubble structure in the wake of a knot has been previously seen in
the case of knot A1 in HST images \citep{Bacciotti2000}. It may thus
be a common feature in DG Tau.
 
We did not include the MV knot identified by \cite{Pyo2003} in our
timescale analysis because the estimated proper motion of
0\farcs28~yr$^{-1}$ yields a ratio of proper motion to radial velocity
60\% higher compared to the HV knot. This is not consistent with
ballistic motions of a gas parcel along the jet axis, which would
predict a constant ratio of proper motion to radial velocity (equal to
tan($i$), with $i$ the jet inclination to the line of sight).
Therefore the MV knot of Pyo et al. does not appear to trace a
separate ejection event, but rather a brightness enhancement following
the HV knot A3.

\subsection{Iron depletion in the blueshifted jet}
\label{sec:depletion}

The dust content of stellar jets is a key parameter in
determining whether the outflowing material originates inside or
outside the dust sublimation radius.  The gas phase depletion of
refractory species such as Fe or Ca, measured with respect to the
solar abundance, offers a direct indication of the presence of dust in
the jet. In previous studies of HH~jets by \cite{Nisini2002, Nisini2005,
Pesenti2004,Podio2006, Podio2009}, relatively high Fe and Ca gas-phase
abundances of 20\%--80\% the solar value were inferred, but most of
these observations probe relatively large distances from the driving
sources $\ge$ 10\arcs, where the original dust content of the flow may
have been modified, eg. through dust destruction in shocks. Exceptions
are \cite{Nisini2005} and \citet{Podio2009} who reported Fe and Ca
depletions at 1--3\arcsec\ of the source in the HH~1 and HH~24 jets,
corresponding to 450--1300 AU at the distance of Orion.  Our
sub-arcsecond resolution SINFONI observations of the DG Tau jet in
\feii\ provide the first opportunity to constrain the dust content of
the blueshifted jet down to only 50 AU (0\farcs3) from its driving
source, i.e. much closer than achieved in previous depletion studies of
HH jets.

Iron depletion is estimated by comparing \feii\ fluxes with lines from
species that show little or no depletion onto grains, such as sulfur
\citep{Beck-Winchatz1996,Pesenti2003}, oxygen \citep{Mouri2000},
hydrogen \citep{Nisini2002}, or phosphorus \citep{Podio2006}. Because
our SINFONI data include no lines from the above species, we use here
previous flux-calibrated \oi$\lambda$6300\AA~observations of the
inner regions of the DG~Tauri blueshifted jet obtained in January 1998 by
\cite{Lavalley-Fouquet2000} with OASIS/CFHT with an angular resolution
of 0\farcs4 and a velocity resolution of 90~\kms.

\begin{figure}
\resizebox{\hsize}{!}{\includegraphics{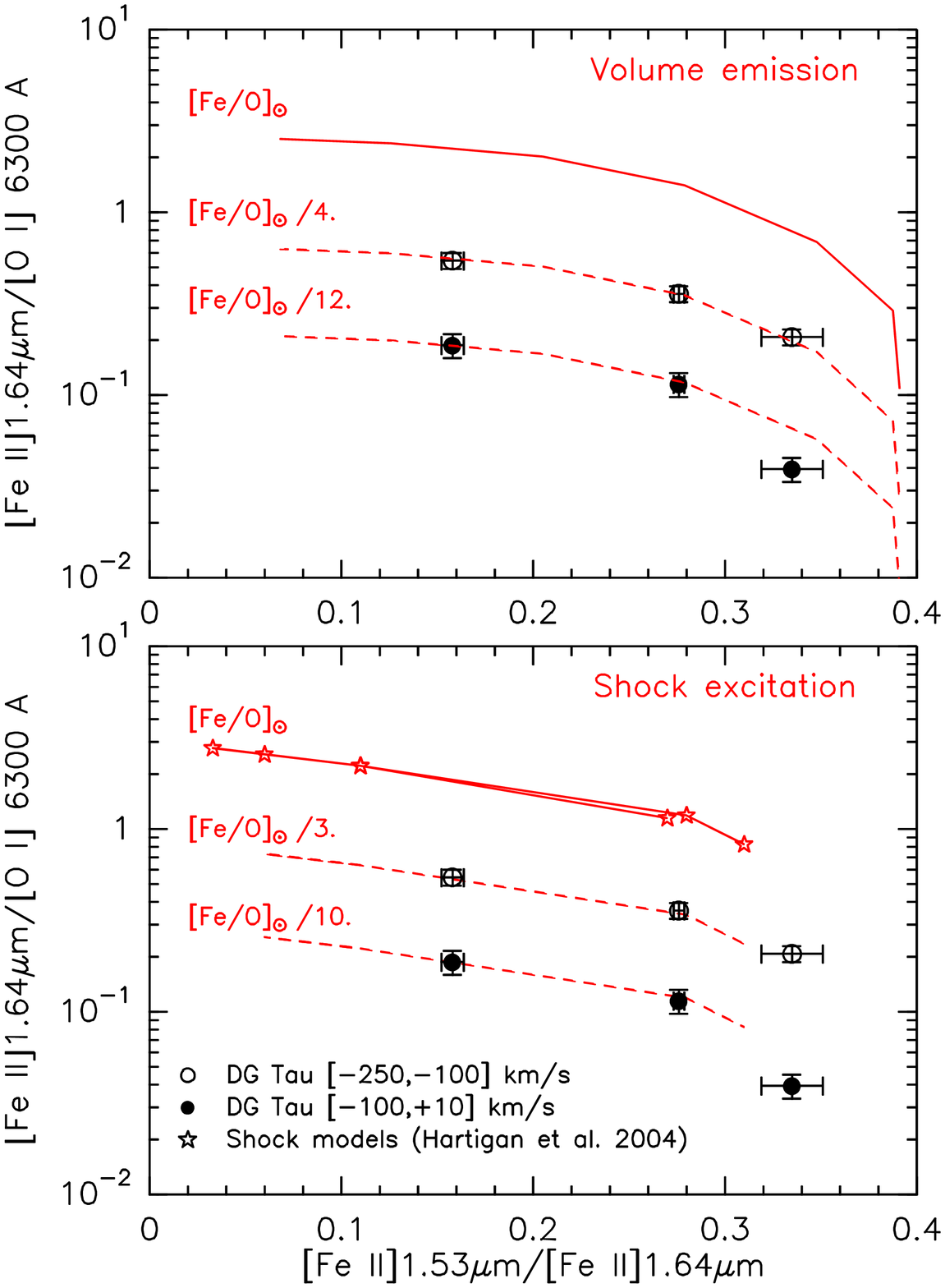}} 
\caption{Line ratio diagnostic diagram for Fe gas phase depletion: \feiil$/$\oi~6300\AA\ versus \feii~1.53/1.64 $\mu$m. \feii\ line fluxes are integrated over the same velocity and spatial intervals as \oi\ fluxes from \cite{Lavalley-Fouquet2000}. The three data points correspond to distances along the DG~Tauri blueshifted jet of 0\farcs25, 0\farcs75 and 1\farcs25 (increasing with decreasing \feii\ 1.53/1.64 $\mu$m). Full curves show predictions for volume emission at 8000~K ({\sl top panel}) and for J-shocks ({\sl bottom panel}) from \cite{Hartigan2004b} assuming an [Fe/O] solar gas phase abundance of 0.062 \citep{Asplund2005}. 
Dashed curves show [Fe/O] abundances required to reproduce observed
line ratios. They indicate Fe gas phase depletion by a factor 3-4 for
the IVB[-250,-100]\kms\ and 10-12 for the LVB[-100,+10]\kms\
intervals.}
\label{fig:depletion}
\end{figure}

We show in Fig.~\ref{fig:depletion} the \feiil$/$\oi~6300\AA~ line
ratios derived from integration of our data over the same velocity
intervals IVB[-250,-100]\kms and LVB[-100,+10]\kms\ and the same
distance ranges along the jet as used by \cite{Lavalley-Fouquet2000}:
d$_{//}$ = -0\farcs225, -0\farcs725, -1\farcs225, with pixel size of
0\farcs5 along the jet. These ratios are plotted against the
1.53\mic/1.64\mic\ ratio, averaged over the same velocity and distance
intervals.
 
In the bottom and top panels of Fig.~\ref{fig:depletion} we compare
there with predictions for a solar abundance ratio Fe/O = 0.062
\citep{Asplund2005} in two excitation cases: (i) a stratified planar shock wave (denoted "shock method" hereafter), and (ii) an emitting volume of homogeneous density and temperature (denoted "volume method" hereafter). 

For atomic shock waves, we considered the models presented in
\cite{Hartigan2004b}, covering preshock densities $n_{\rm H} =
10^3-10^5$ cm$^{-3}$, shock speeds of 30--50 \kms, a preshock
ionization fraction of 1\%-50\%, and a preshock magnetic field of
30--200$\mu$G.  We find that the observed \feii$/$\oi\ line ratios in
the DG Tau jet are systematically lower than the predictions for a
solar Fe/O ratio, by a factor 3 in the IVB interval and by a factor 10
in the LVB interval, independent of the distance from the source. Note
that the predicted line ratio curve assumes that the mass-flux
entering the shock was the same in 1998 (\oi\ data) and 2005 (\feii\
data), i.e. that the twice higher electronic density in \feii\ in 2005
compared to \sii\ in 1998 (Section~\ref{sec:ne}) is mostly caused by
ionization gradients within the shock cooling zone. This gives a
conservative lower limit to iron depletion. If, instead, the increase
in \ne\ between the \oi\ and \feii\ observations were caused by an
increase in pre-shock density, one would predict a higher
\feii$/$\oi\ ratio and infer a larger iron depletion in the jet.
 
For the uniform volume method, assumptions need to be made on the
temperature and ionization state of the gas: we assume a temperature
of 8000~K typical of stellar jets, that iron is all in the form of
Fe$^+$ and oxygen is mostly neutral (as expected at such
temperatures). The predicted \feii$/$\oi\ ratio assumes that the mass
in each element of volume was about the same between 1998 and 2005,
i.e. that the difference in \ne\ between the two data sets is caused by
an increase in ionization fraction between the two epochs. The results
are remarkably similar to the shock method: an Fe/O ratio lower
than solar by a factor 4 in the IVB interval and a factor 12 in the
LVB interval. Again, inferred depletions would increase if the total
jet density $n_H$ (instead of \xe) increased between the two
epochs. Finally, we note that applying an extinction correction to the
observed line fluxes would decrease the observed \feii$/$\oi\ ratio,
again increasing the inferred depletion in both methods.

To evaluate a possible effect of time variability in the \oi\ line, we
compared the \oi\ fluxes measured in 1998 by
\cite{Lavalley-Fouquet2000} against the flux-calibrated HST/STIS
transverse PV diagram obtained on 2003 December 1 at a distance of
-0\farcs3 from the star \citep{Coffey2007}. The \oi\ line fluxes per
arcsecond length of the jet in the IVB and LVB intervals in the latter
data ($1.9 \times 10^{-13}$ and $1.3 \times 10^{-13}$ ${\rm erg s^{-1}
cm^{-2} arcsec^{-1}}$ respectively; Coffey, private communication) are
80\% and 5\% higher than their respective values in December 1998 at the
same distance, averaged over 0\farcs5 along the jet.  Using the Coffey
et al. \oi\ fluxes would thus increase depletion in the IVB range at
0\farcs3 by a factor 2, while depletion in the LVB would remain
unchanged.  However, line fluxes measured in  the narrow HST slit
(of width 0\farcs1) will also be more affected by the precise
position of knots at the epoch of observation than fluxes averaged
over 0\farcs5 along the jet. Results obtained by comparison with
Lavalley \oi\ data may thus be more reliable despite the longer time
span.

We thus tentatively interpret the low \feii$/$\oi\ ratios observed as
indication of iron gas-phase depletion by a factor $\simeq 3-4$ in
gas faster than -100 \kms\ and by a factor of $10-12$ at speeds below
-100 \kms. This would correspond to about 33\% and 10\% of elemental
iron in the gas phase, respectively.  Simultaneous observations of the
DG Tau jet involving refractory and non-refractory species would be
extremely valuable to confirm the present findings.  The observed
trend of higher depletion at lower flow velocities, also seen in
\cite{Podio2009} in calcium toward the HH~111 jet, would imply that
slower jet material in the jet originates in regions farther from
the star than higher velocity material, or that it has been less 
shock-processed.  Implications for launching models are discussed in
Section~\ref{sec:models}.

\subsection{Blueshifted jet mass-flux in the \feii\ emitting components}
\label {sec:mdot}

From our \feii\ observations we derive estimates of the mass-flux
rates in the atomic component of the blueshifted jet as a function of
the distance from the star, and for the two velocity intervals HVB and
MVB. We used three different methods, the first one based on estimates
of the jet density and cross-section, the other two based on the
\feii\ luminosity, corrected for depletion. We detail below the
assumptions and parameters for these three methods. The resulting
mass-fluxes are plotted as a function of distance in
Figure~\ref{fig:mass-loss}.

\subsubsection{Method 1: from jet density and cross-section}

For a jet of cross-section $A_J$ and velocity $V_J$, uniformly filled with a density of hydrogen nuclei $n_H$, the mass-flux rate is given by
\begin{displaymath}
\dot{M}_J = \mu m_H n_H A_J V_J,
\end{displaymath}
where $\mu$, the average weight per hydrogen nucleus, is $1.4$ for a cosmic gas with 10\% helium. Assuming a circular cross section of diameter $D_J$, the above equation can be re-written as
\begin{equation}
\dot{M}_J = 1.3 \times 10^{-9} M_{\odot} yr^{-1} \left ( \frac{n_H}{10^5~{\rm cm}^{-3}} \right ) \left ( \frac{D_J}{14~AU} \right)^2 \left ( \frac{V_J}{100~{\rm km s}^{-1}} \right ).
\label{eq:massloss1}
\end{equation}
This method has the advantage of being unaffected by iron depletion or
by circumstellar extinction.  We plot in Figure~\ref{fig:mass-loss}
(dashed-dotted line) the mass-loss rates derived from this method and
computed every 0\farcs1 along the blueshifted jet.

We used an average deprojected velocity $V_J = 300$ \kms\ for the HVB
and $V_J = 135$ \kms\ for the MVB ranges, derived from the average
radial velocities of $V_r \simeq$ 220 and 100 \kms\ of the spectral
components resolved by \citet{Pyo2003}, with a jet inclination angle
with respect to the line of sight of i=42$^{\circ}$
\cite{Lavalley2000PhD}. To estimate the HVB cross-section, we took
$D_{HVB}^2 = FWHM_{HVB}^2-FWHM_{0}^2$ where FWHM$_{HVB}$ is the
transverse \feii\ emission width measured in the HVB deconvolved maps,
and FWHM$_{0} = 0\farcs05-0\farcs08$ is the effective resolution of
these maps. To estimate the cross-section of the MVB flow, we subtracted
the cross-section of the central HVB component nested inside it,
i.e. we take $D_{MVB}^2 = FWHM_{MVB}^2-FWHM_{HVB}^2$. The effective
resolution of the deconvolved maps cancels out in this case.
 
Estimates of the total density n$_H$ along the jet are derived from
\ne /\xe, where the electronic densities \ne\ along the jet are
obtained from our \feii 1.533/1.644 line ratios. For the hydrogen
ionization fraction \xe, we used linear fits to the values derived by
\citet{Bacciotti2002} in 1999 with HST/STIS from optical line ratios
on similar spatial scales and velocity ranges.  One potential
limitation in using these values may come from time
variability. However, the \xe\ values obtained in 2003 at $d$ =
-0\farcs3 by \citet{Coffey2008} agree very well with the
1999 values of \citet{Bacciotti2002}, indicating moderate variability
on a three year timescale in this parameter.
Our mass-flux estimates at $d = -0\farcs3$ in the HVB and MVB
intervals differ slightly from those obtained by \citet{Coffey2008}
using the same method (star symbols in Figure~\ref{fig:mass-loss}), 
the discrepancy mainly comes on the one hand from a jet radius twice as large as
our estimate (they do not correct for instrumental PSF) and because
of the quadratic dependence, this results in a factor 4. On the other
hand, the total density obtained by \citet{Coffey2008} is a factor 3
lower than ours, which compensates the factor 4 from the jet width. The
difference in  MVB  also arises because the authors do
not correct the MVB width by the area filled by the HVB, which we do.
 
\subsubsection{Method 2: from \feii\ luminosity, volume emission from uniform jet slab}

The second method derives the mass-loss rate from \feiil\ luminosities
assuming optically thin volume emission from a uniform jet slab. This
method, described in detail in \cite{Hartigan1995} for optical lines,
assumes uniform \Te, \xe\ and \ne\ and relates the line flux to the
emitting mass within the elementary aperture. Using a 16-level
excitation model of Fe$^+$ \citep{Pesenti2003}, we fitted the emissivity
of the 1.64$\mu$m line as a function of \ne\ for $T_e \simeq 10^4$~K
as
\begin{eqnarray}
j(1.64\mu{\rm m}) =  2 \times 10^{-17} \times \left ( 1+ \frac{3.5\times10^4}{n_{\rm e}({\rm cm^{-3}})} \right )^{-1}  {\rm erg~s^{-1}~sr^{-1}~ion^{-1}}
\label{eq:emiss164}
\end{eqnarray}
(a good approximation for \ne\ $\ge 3000$ \cm). The mass-loss rate is then given by the mass divided by the pixel crossing time
\begin{eqnarray}
\nonumber
\dot{M}_J & =  & 1.45 \times 10^{-8}~M_{\odot} yr^{-1} \left ( 1+ \frac{3.5\times10^4}{n_{\rm e}(cm^{-3})} \right )  \left ( \frac{L_{\rm [Fe II]}}{10^{-4} L_{\odot}} \right ) \\ 
& & \times \left ( \frac{V_t}{150~km~s^{-1}} \right ) \left ( \frac{l_t}{2\times10^{15}~cm}\right )^{-1} \left( \frac{\rm [Fe/H]}{\rm [Fe/H]_{\odot}} \right)^{-1},
\label{eq:massloss2}
\end{eqnarray}
with $l_t$ the size in the plane of sky of the elementary aperture along the jet axis, $V_t$ the tangential flow velocity in the plane of sky, and
$\rm [Fe/H]_{\odot} = 2.82 \times 10^{-5}$ \citep{Asplund2005}. 

We computed the \feiil\ luminosity in the HVB and MVB channel maps in
rectangular apertures of projected length along the jet axis
$l_t=0.1\arcsec=2\times10^{14}$~cm (at 140pc), and width $\pm
0\farcs5$ across the jet. We adopted the same values of \ne\ as a
function of distance as in method~1 and the same radial velocities to
derive $V_t = V_r \times \tan(i)$. Finally, we applied the correction
for the Fe gas-phase depletion estimated in the previous section,
using the "volume" hypothesis, ${\rm [Fe/H]}/{\rm [Fe/H]_{\odot}}=$
1/4 and 1/12 for the HVB and MVB respectively. The mass-loss rates
derived from method 2 are plotted in full line in
Figure~\ref{fig:mass-loss}.

\subsubsection{Method 3: from \feii\ luminosity, emission from shock fronts}

The third method to estimate the jet mass-loss rate from the \feii\
luminosity assumes that the line emission in the elementary aperture
is arising from shock fronts.  It is described in detail in
\cite{Hartigan1995} for the \oi\ line. The mass-flux entering the
shock, $\dot{M}_s$, is given by
\begin{displaymath}
\dot{M}_S = \mu m_H n_H V_s A_s,
\end{displaymath}
with $\mu$ = 1.4, and $V_s$ and $A_s$, the shock velocity and cross-section respectively. 

From the planar atomic J-shock models of \cite{Hartigan2004b}, we derive a proportional relation between $n_H V_s$ 
and the emitted \feiil\ line flux per shock area (multiplied by 2 to account for both sides of the shock)
\begin{displaymath}
F_{\rm 1.64\mu m} =  9.69\times 10^{-14}\times n_H V_s  \left (   \frac{\rm [Fe/H]} {\rm [Fe/H]_{\odot}} \right ) {\rm erg~s^{-1}~cm^{-2}},
\end{displaymath}
where we scaled the model fluxes to the recent estimate of solar  abundance $\rm [Fe/H]_{\odot} = 2.82 \times 10^{-5}$ \citep{Asplund2005}. 
Integrating over the shock area $A_s$ yields the following equation for the shock mass-flux
\begin{equation}
\dot{M}_S = 9 \times 10^{-8} M_{\odot} yr^{-1} \left ( \frac{L_{\rm [Fe II]}}{10^{-4} L_{\odot}} \right ) \left (   \frac{\rm [Fe/H]} {\rm [Fe/H]_{\odot}} \right )^{-1}. 
\end{equation}

The jet mass flux is related to the mass-flux entering the shock(s) inside the aperture, $\dot{M}_s$, by the following equation \citep[see][]{Hartigan1995,Cabrit2002}
\begin{equation}
\dot{M}_J = \dot{M}_s \left ( \frac{V_J}{V_s} \right )  \left ( \frac{\cos\theta}{N_{shock}}\right ),   
\label{eq:massloss3}
\end{equation}
where cos($\theta$) = $A_J/A_s$ for oblique shocks at an angle $\theta$ with the flow normal, and $N_{shock}$ is the number of shock fronts within the integration aperture. 

Here we assume perpendicular ($cos\theta =1$) shock fronts located
every 50~AU along the jet (maximum cooling length in the models of
\cite{Hartigan2004b}). This will tend to give an upper limit to the mass-flux.The \feii\
luminosity in the HVB and MVB channel maps is integrated every
0.3\arcsec\ = 50 AU along the jet axis, so that $N_{shock} = 1$. In
addition, following the estimates of shock velocities derived by
\cite{Lavalley-Fouquet2000}, we adopted $V_{J}/V_{s}=4$ in the HVB and
$V_{J}/V_{s}=2$ in the MVB intervals. Finally, we corrected for the Fe
gas-phase depletion estimated in the previous section in the shock
hypothesis, using ${\rm [Fe/H]} / {\rm [Fe/H]_{\odot}}=$ 1/3 and 1/10
for the HVB and MVB respectively. The derived mass-loss rates are
represented with a dashed line in Figure~\ref{fig:mass-loss}.

\subsubsection{Comparison of the three methods}

\begin{figure}
\includegraphics[width=0.5\textwidth]{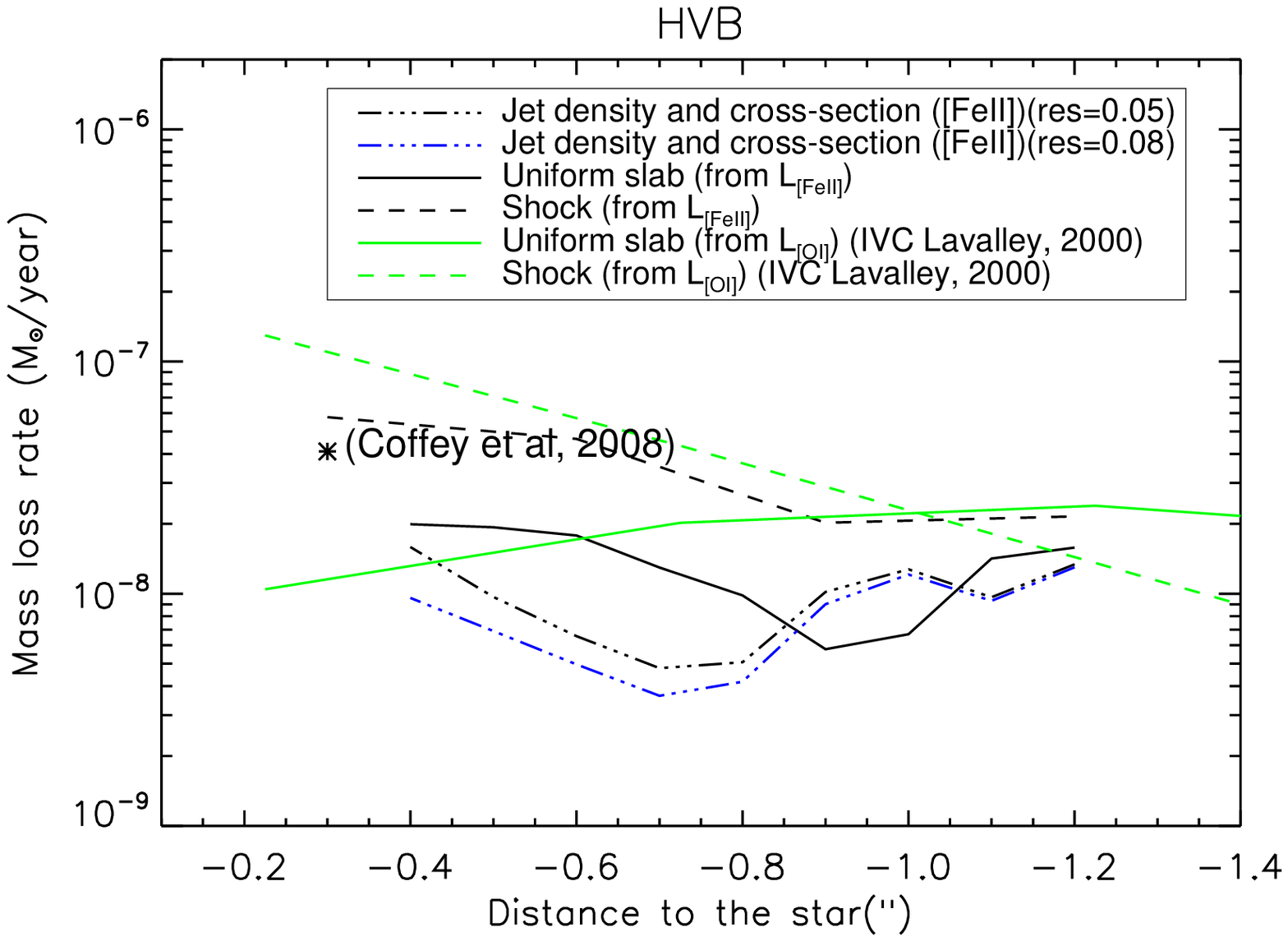} 
\includegraphics[width=0.5\textwidth]{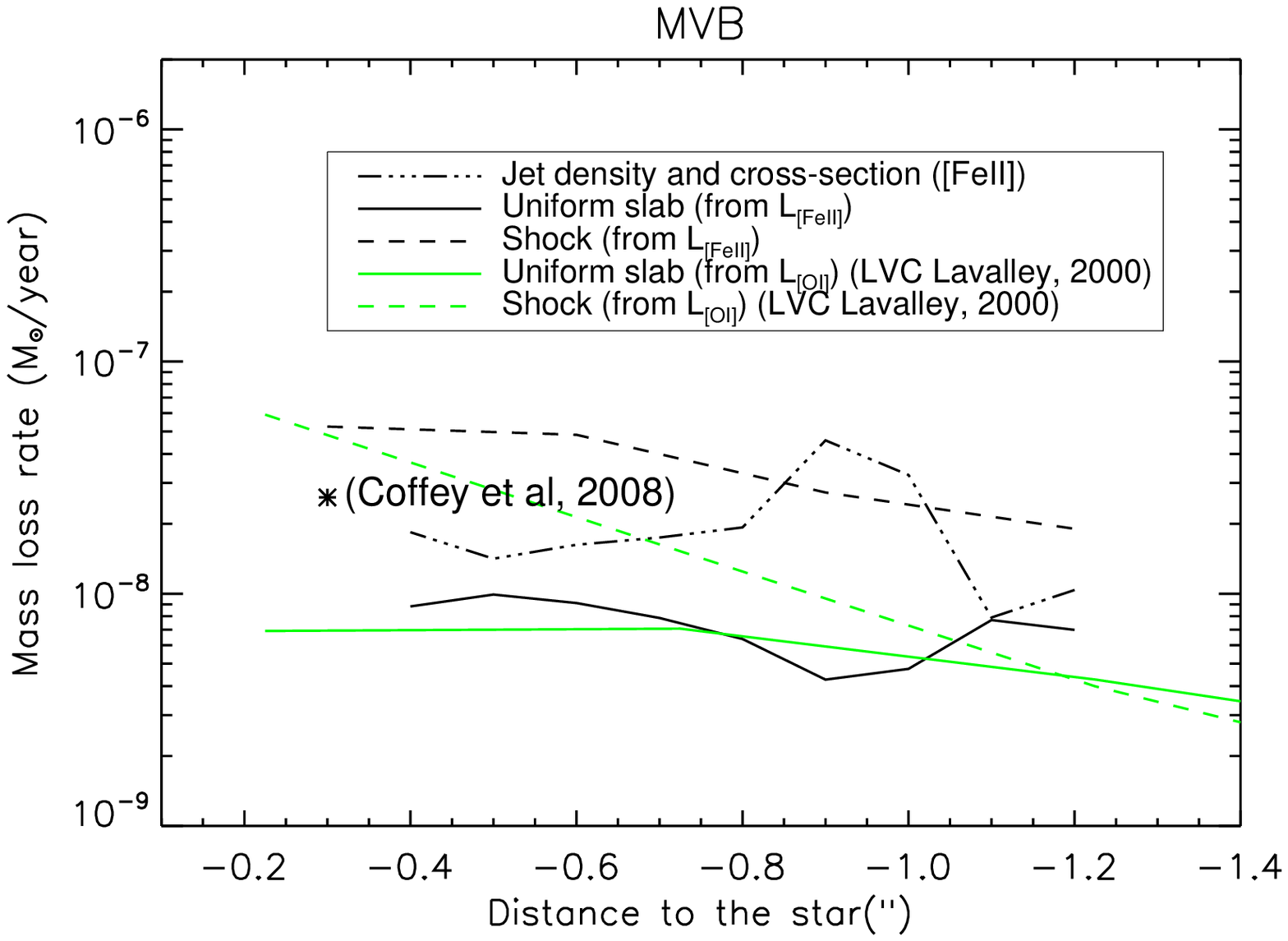} 
\caption{Values of the mass-loss rate for the atomic blueshifted jet as a function of projected distance to the star and for the two velocity components: HVB ({\sl top panel}) and MVB ({\sl bottom panel}).
The values derived from the \feiil\ observations using the three methods discussed in the text are plotted with black curves: jet density and cross-section (dashed-dotted curve), uniform slab (full curve), shock front (dashed curve). The values obtained by \cite{Lavalley-Fouquet2000} using the optical line \oi\  and methods 2 and 3 are shown for comparison (gray/green curves). 
Star symbols show the values obtained by \cite{Coffey2008} from an optical HST/STIS transverse slit at -0\farcs3.}
\label{fig:mass-loss}
\end{figure}

We plot in Figure~\ref{fig:mass-loss} the mass-loss rate estimates in the blue DG Tau jet  as a function of distance to the star obtained for the three methods, with a separate panel for the HVB [-300,-160]  \kms\ and MVB [-160,-50] \kms velocity intervals. 
The three methods agree reasonably well within a factor 2-5 at all distances from the star, lending some confidence to these values, and 
to our depletion estimates. 

For comparison, the mass-fluxes obtained by
\cite{Lavalley-Fouquet2000} in their IVB and LVB intervals using the
optical line \oi\ and methods 2 and 3 are also shown in
Figure~\ref{fig:mass-loss} (gray/green curves). They were scaled up by
1.44 to be consistent with the new solar oxygen abundance [O/H]$_\odot
= 4.57 \times 10^{-4}$ of
\citet{Asplund2005}. Because the iron depletion was inferred from the \oi\ to \feii\ luminosity ratio assuming the same jet density in the two epochs, it is expected that
both lines should give identical mass-fluxes. The slight differences
observed stem from the use of slightly different velocity intervals
(IVB, LVB rather than HVB, MVB), and from the fact that the \oi\
luminosity is not strictly proportional to $n_H V_s$ in shocks (the
slope is steeper than 1), so that method 3 is less accurate for \oi\
than for \feii.

Averaging our \feii\ mass-flux estimates over the central 1\arcsec\
gives a mass-loss rate of $1.7 \pm 0.7 \times 10^{-8} M_{\odot}
yr^{-1}$ for the HVB and $1.6 \pm 0.9 \times 10^{-8} M_{\odot}
yr^{-1}$ for the MVB using the three methods (the quoted uncertainty range
is the 1-sigma dispersion about the mean). Our derived HVB
mass-loss rate is consistent with the lower limit of $1.2 \times
10^{-8} M_{\odot} yr^{-1}$ estimated by \cite{Beck2010} from
high-velocity extended HI Br$\gamma$ emission. The total mass-loss
rate for the \feii\ emitting atomic gas in the velocity range -300 to
-50 \kms\ is therefore $\dot{M}_{J} \simeq 3.3 \pm 1.1 \times 10^{-8}
M_{\odot} yr^{-1}$, with approximately equal contribution from the HVB
and the MVB. This value is a lower limit to the total mass-loss rate
of the atomic component because the \feii\ emission does not probe the whole
range of velocities seen at optical wavelengths.
 
\subsection{Accretion rate and ejection-to-accretion ratio in DG Tau}
\label{sec:accretion}

\cite{Muzerolle1998b} have shown that the luminosity of the Br$\gamma$
line in a large sample of T Tauri stars is well correlated with the
accretion luminosity onto the star inferred from modeling of the
optical-UV excess. During our SINFONI run, we also obtained K-band
data including the Br$\gamma$ line.  We find that any spatially
Br$\gamma$ extended component contributes at most 2 \% of the total
line flux integrated over all velocities. This is consistent
with the recent finding of \cite{Beck2010}, who show that
although extended high-velocity Br$\gamma$ emission is detected along
the DG~Tau blueshifted jet, the extended component makes up 2 \% of
the total (velocity integrated) line flux. We therefore make the
assumption below that all the HI Br$\gamma$ flux in
DG~Tau is related to the accretion process.  From our flux-calibrated
data we derived a K-band continuum of 7.1 mag and a Br$\gamma$
equivalent width of 5.6~\AA\ corresponding to a line brightness of 
$3.9 \times 10^{-16}$ W~m$^{-2}$ and line luminosity of
$2.5 \times 10^{-4}$ \Lsun\ after extinction correction by
$A_V=1.6$mag \citep{Gullbring2000}. Using the empirical fit of
\cite{Muzerolle1998b}, we derived an accretion luminosity L$_{\rm acc}$
= 0.77L$_{\sun}$ for DG Tau in October 2005. Assuming that the luminosity
arises from an accretion shock at the base of an accretion column
anchored at the inner disk truncation radius $R_{i}$, the mass
accretion rate onto the star is then obtained from
\citep{Gullbring1998}

\begin{equation}
\label{eq:accretion}
\dot{M}_{\rm acc}= \frac{L_{\rm acc}R_\star}{GM_\star} \Big( 1-\frac{R_\star}{R_{i}} \Big)^{-1} .
\end{equation}  

Taking $R_{i} \simeq 5 R_\star$ as in \cite{Gullbring1998}, and
adopting for DG~Tau a stellar radius of $R_\star$=2.3 R$_{\sun}$ and a
stellar mass of $M_\star$=0.7 \Msun \citep{Hartigan1995} we derived an
accretion rate of $\dot{M}_{\rm acc} = 10^{-7}$ \Msun\yr. With a
similar Br$\gamma$ flux (4.5 $\times$ 10$^{-16}$ W m$^{-2}$,
\cite{Beck2010} derive $\dot{M}_{\rm acc}$ =
$9.6~\times~10^{-8}$~\Msun\yr using A$_V$=3.3~mag and a stellar mass
of 2.2~\Msun for DG~Tau from \cite{White2004}. We note that this
value of \macc\ only probes the {\it instantaneous} accretion rate
from the inner disc onto the stellar surface in October 2005. It is not
necessarily representative of the mean accretion level pertaining
to the period 2002-2004 when the jet material probed by our SINFONI
observations was ejected (see Fig.~\ref{fig:knots}).

Indeed, diagnostics of the accretion shock, such as the Br$\gamma$ and
Ca II line strength or the optical veiling, are known to be highly
variable in T Tauri stars on scales from days to years, especially in
stars as active as DG Tau. \cite{Muzerolle1998b} measured in 1998 a
Br$\gamma$ equivalent width 3 times larger than ours,
EW(Br$\gamma$)=14.1~\AA, and obtained a 3 times higher accretion rate,
assuming a typical K-band magnitude for DG Tau K= 6.74 (average of 17
measurements, provided in \citep{Kenyon1995}).
\cite{Gullbring2000} derived $\dot{M}_{\rm acc} = 5 \times 10^{-7}$ \Msun\yr, by fitting an accretion shock model to the Balmer and UV excess continuum emission measured in 1996. Finally,  the mean correlation between CaII line luminosity and accretion rate in T Tauri stars suggests $\dot{M}_{\rm acc} \simeq 2 \times 10^{-7}$ \Msun\yr\  in DG Tau in 2002-2003 \citep{Mohanty2005}.  

Below we therefore adopt $\dot{M}_{acc} = (3 \pm 2) \times 10^{-7}$
\Msun\yr\ as the possible range of accretion rate prevailing in
DG~Tau. This is significantly lower than the $\dot{M}_{acc} = 2 \times
10^{-6}$ \Msun\yr\ obtained by \cite{Hartigan1995} from an analysis of
the optical veiling in 1988-1989.  However, the method of HEG95
systematically tends to overestimate accretion rates in T Tauri stars
by an order of magnitude on average \citep{Gullbring1998}.

From the range of mass-flux rates derived in the \feii\ jet (see
Section~\ref{sec:mdot}), we thus inferred  for the blueshifted jet
lobe a ratio of ejection to accretion rates of 0.04--0.4 for the
\feii\ emitting flow, with about equal contributions from the MVB and
HVB ranges. Most of the uncertainty in this ratio (a factor 5) arises
from the mass-accretion rate.  The ratio is substantially higher than
previously reported by \cite{Lavalley-Fouquet2000}, mainly because
they adopted a 10 times higher accretion rate from
\cite{Hartigan1995}. It is compatible with the average  ejection to
accretion ratio for one jet lobe of $\simeq 0.1$ in atomic T Tauri
jets obtained by \citet{cabrit-iau243} when comparing \oi\ mass-flux
rates from spatially unresolved data \citep{Hartigan1995} with revised
mass-accretion rates from \citet{Gullbring1998}. Our mass-flux
estimates therefore do not appear to suffer from a strong systematic bias.
 
\subsection{Implications for jet launching models}
\label{sec:models}

Our results on mass-fluxes, kinematics, and iron depletion in the DG Tau jet  have several implications for proposed jet origins. 

First, we note that the observed velocities in excess of $-50~$ \kms\ in the
blueshifted \feii\ flow are much too high to be produced by disk
photo-evaporation. Indeed, in photo-evaporated flows terminal
velocities reach at most 3-4 $\times$ the sound speed $c_S$ at the
base of the flow \citep{Font2004}. Maximum temperatures of $\simeq$
$10^{4}$~K reached at the inner disk surface through EUV and X-ray
irradiation \citep{Ercolano2009, Gorti2009} would thus lead to terminal
velocities $\le$ 30-40~km~s$^{-1}$, much lower than observed in \feii.

 One might wonder if the MVB flow could trace matter entrained and
 accelerated in a mixing layer between the high-velocity jet (traced
 by the HVB) and the outer photo-evaporated disk wind.  The most
 recent computations of disk photo-evaporation \citep{Owen2010,
 Gorti2009}, including X-ray irradiation, indeed predict mass-flux
 rates that could reach 10$^{-9}$-10$^{-8}$ M$_{\odot}$ yr$^{-1}$ over
 the entire disk, sufficient to supply the MVB flux. However, the same
 models show that most of the photoevaporated mass-flux arises from
 disk radii $r \ge$ 5~AU \citep{Gorti2009,Owen2010}.
The corresponding streamlines \citep[see][Fig.~3]{Owen2010} have cylindrical radii $>$~20~AU for altitudes above the disc z~$>$~30~AU, significantly wider than the derived HVB jet radius (see
Figure~\ref{fig:jetwidths}); therefore, one expects very little or no
entrainment of the photoevaporated flow by the HVB jet.
 
Second, we note that the gas-phase depletion of iron, if confirmed,
rules out an origin in the stellar surface, at least for the \feii\
MVB emitting material. Indeed, recent determination of
photospheric [Fe/H] for young stars in Taurus are compatible with
solar abundance \citep{Santos2008,dOrazi2011}. Lower [Fe/H] ratios
($\simeq$ 0.2 solar) have been derived in X-rays tracing coronal
material but in these observations the [Fe/O] abundance ratio is
0.4-0.8 solar \citep{Scelsi2007,Guenther2007}. These abundance patterns
cannot reproduce the low [Fe/O] ratios ($\simeq$ 0.1 solar) inferred
in the MVB emitting gas, indicating that the observed Fe gas phase
depletion is likely caused by depletion onto dust grains. Our observed
Fe depletion pattern also excludes the fact that the
medium-velocity conical sheath surrounding the jet could trace a
cocoon of matter ejected sideways from the narrow high-velocity
beam.  Sideways ejected material should be {\it equally or less}
depleted onto dust (more shock-processed) than the faster
central beam, whereas we observe the opposite, with a 3 times larger
depletion at lower velocity (see Section~\ref{sec:depletion}).  It is
difficult to pinpoint more precisely the launch radius from the
depletion alone; the dust sublimation radius in DG Tau appears to be
$r_{\rm sub} \simeq$ 0.14 AU from K-band interferometric measurements
\citep{Akeson2005}. However, the K-band size may be only an upper
limit to $r_{\rm sub}$ owing to the effect of scattered light
\citep{Pinte2008}.  Ejection from near corotation ($r_{\rm corot} =
0.06$~AU in DG Tau with a stellar period of 6.2 days and assuming a
0.7\Msun\ star) is thus not totally excluded at this stage, pending
more detailed models. Ejection from slightly beyond 0.1 AU is not
ruled out either, because sublimation starts further out for smaller
grains, and only 10\%-33\% of the dust appears to have been destroyed
in the \feii\ jet. In addition, $r_{\rm sub}$ could be larger than the
K-band radius if the inner disk of DG Tau is optically thick in the
gas continuum.
 
Below, we confront the DG Tau jet properties with predictions for
proposed ejection processes, including the X-wind model and an
extended MHD disk-wind model.

\subsubsection{The X-wind model}

The X-wind model assumes that the disk is not intrinsically magnetized
and is threaded only by the stellar dipole. It posits that stellar
fieldlines are bunched up at the corotation radius and are partly opened
in a fan-like configuration along which a centrifugal wind, carrying
off the angular momentum from the funnel flow, is assumed to be
launched. The semi-analytical wind solution is characterized by an
axial density enhancement (the jet) surrounded by a wide-angle
flow, where the speed is nearly constant across all streamlines
\citep{Shang1998}. This predicted constant flow speed stems from the
cylindrical shape of the Alfv\'en surface, which appears to be a
generic property of the X-wind \citep{Cai2008}. Because all streamlines
are launched from the same radius $r_0 = r_{\rm corot}$, their
magnetic lever arm parameter $\lambda = (r_A/r_0)^2$ is also the same,
and so is the terminal speed after magneto-centrifugal acceleration
\citep{BP82}
\begin{equation}
V_p^\infty = \sqrt{2\lambda-3} V_{Kep}(r_0). 
\label{eq:vterm}
\end{equation}
The steep velocity decrease across the DG Tau blueshifted jet seen in
Figs~4 \& 5, from V=-200 \kms in the HVB for angles to the jet axis
less than 4$^{\circ}$ to V=-100 \kms in the MVB for angles to the jet
axis of 14$^{\circ}$ and down to -30,-40 \kms\ at wider angles in
\oi\ and \sii\ \citep{Bacciotti2002, Coffey2007}, is clearly not
compatible with this characteristic property.  The X-wind model could
account for the \feii\ HVB component only, but this would require that
the MVB flow originates in an MHD disk wind (because it cannot be
explained by photo-evaporation, see above). However, as
discussed in \cite{Ferreira2006}, it is very unlikely that an X-wind
and an extended MHD disc-wind will coexist because this would imply a
hole in the disc magnetic flux distribution between these two
components.  Therefore the X-wind model as currently envisioned does
not explain the kinematics in the \feii\ DG Tau jet.

\subsubsection{Conical wind from the disk-magnetosphere boundary}

Recent numerical simulations of the interaction between a stellar
dipole and a disk with no net magnetic field yield a situation
somewhat different from the (semi-analytical) X-wind model
\citep{Romanova2009}. As the stellar fieldlines are bunched up by
accretion and stretched, a narrow conical wind is formed along the
neutral line (rather than in the angular sector below it), and powered
by the magnetic pressure force (rather than by the centrifugal force
as in the X-wind). In the standard (non-propeller) regime relevant
to DG Tau, where the star rotates slowly, the total mass-flux rate in
the conical wind is about 10\%-30\% of the accretion rate onto the
star, compatible with the observational constraint.  A strong
transverse velocity gradient is also present, with faster matter from
the stellar magnetosphere on the inside and slow matter from the disk
on the outside. However, most of the wind mass reaches a low poloidal
speed of about 40 \kms\ for the CTTS dimensional scalings appropriate
in DG Tau: $v_p \simeq 0.2v_0$ with $v_0 = 197$\kms\ in Table~1 of
\cite{Romanova2009}.  And the half-opening angle of 30\degr--40\degr\
is much wider than the 4--15\degr\ observed in \feii. Therefore the
conical wind as modeled in this work appears both too slow and too
broad to explain the \feii\ jet observations, although it could match
the LV sheath surrounding the jet observed in optical lines
\citep{Bacciotti2002}.  In this framework, the whole \feii\ jet
(HVB and MVB) would have to originate mainly in the stellar
surface. This would contradict with the observed iron
gas-phase depletion in the DG Tau jet, indicative of a substantial
dust content.

\subsubsection{Extended MHD disk wind}

Protostellar disks may indeed harbor a substantial magnetic flux,
advected during the initial infall phase of star formation.  If the
field is slightly below equipartition, rotation remains
quasi-Keplerian in the disk midplane and angular momentum can be
efficiently extracted by magnetic torques to launch a powerful wind
from the disk surface, as shown by analytical and numerical 2D
solutions of the vertical disk structure
\citep{Ferreira1997,Zanni2007}. \citet{Pesenti2004} found that a
self-similar, steady MHD wind from a warm disk \citep{Casse2000},
with a lever arm parameter $\lambda = (r_A/r_0)^2 \simeq 13$ could fit
the tentative jet rotation signatures observed in the low-velocity
\oi\ flow \citep{Bacciotti2002}. The same type of model can also
reproduce observed jet widths in T Tauri stars \citep{Ray2007}. We now
check whether this model could also explain the speed, mass-flux, and
depletion in each of the \feii\ velocity components.
 
We may use the range of flow speeds in each \feii\ velocity component
to constrain the range of ejection radii in the considered disk wind
model.  The asymptotic velocity given in Eq.~\ref{eq:vterm} is
reached only far from the star in self-similar solutions
\citep{Ferreira2006}.  We therefore use here the exact predicted flow
speeds at an altitude of 50~AU above the disk for the $\lambda = 13$
model, plotted as a thick black curve in Fig.~3 of
\citet{Ferreira2006} (note that velocities in that graph are scaled by
$1/\sqrt{M_\star/M_\odot}$). The radial velocity range of 160 to 250
\kms\ for the HVB component \citep{Pyo2003} corresponds to deprojected
flow speeds of 210--350 \kms and to launch radii of 0.08 AU to 0.25
AU for a stellar mass ${M_\star} = 0.7M_\odot$. The upper value is
consistent with the outer launch radius of 0.2--0.5 AU for the HVB gas
derived from tentative rotation signatures in optical/near-UV lines
\citep{Coffey2007}. Similarly, the radial velocity range of [-50,-160]
\kms\ of the MVB, and deprojected speeds of 70--210 \kms, would
imply launch radii of 0.25--1.5~AU.  The observed Fe depletions could
be explained if 10\% of dust is already destroyed around 1 AU and 33\%
around 0.2 AU. Detailed modeling including a dust size distribution
and opacity of the gas disk are needed to see whether this is
compatible with K-band size measurements \citep{Akeson2005}.

The ejection-to-accretion ratio for one jet lobe in a self-similar MHD disk wind powered by angular momentum extraction from the disk is given by \citep{Ferreira2006}
\begin{equation}
{{\dot M}_{J}\over {\dot M}_{\rm acc}} = {1 \over  4(\lambda-1)} \log(r_e/r_i)  = {1 \over  2(\lambda-1)} \log(V_{\rm max}/V_{\rm min}) ,
\label{eq:mejmacc}
\end{equation}
where $r_e$ and $r_i$ are the external and internal radii of the wind
launch zone, and $V_{\rm min}$ and $V_{\rm max}$ the corresponding
minimum and maximum flow speeds. With $\lambda=13$, and values of
$V_{\rm min}$ and $V_{\rm max}$ as above, one expects an
ejection/accretion ratio for one jet lobe of 0.04 in the MVB and 0.02
in the HVB intervals, respectively. This is compatible within
uncertainties with the mass-fluxes estimated here (see
Section~\ref{sec:mdot}) if the DG Tau accretion rate during the period
(2002-2004) was $ \simeq 5 \times 10^{-7}$ \Msun\yr, near the upper
end of the range reported in recent works (see
Section~\ref{sec:accretion}). Clearly, a closer monitoring of
accretion and ejection signatures in DG~Tau over a few years is
required to estimate more accurately the ejection/accretion ratio in
the DG Tau jet, and provide a more definitive test.

We stress that the above comparison must be taken as indicative
only. First, we have discussed the implications in the context of
a specific disk wind solution with $\lambda$=13. A smaller $\lambda$
solution may be more compatible with the mass flux constraints.  A
thorough investigation of possible disk wind solutions compatible
with all observational constraints should be conducted.  
Secondly, the above comparison assumes that the periodic knots/bubbles do not
significantly modify the overall steady-state disk wind kinematics and
mass-flux (even though they will locally modify its {\it ionization
level} through shocks). The period of knot generation of 2.5 years in
DG Tau is about 140 times the star rotation period of 6.2 days,
i.e. much longer than the orbital timescale in the inner disk, but
comparable to the orbital period at 1.5~AU. Therefore, the
steady-state disk wind assumption could be valid for the HVB,
 but only marginal for the MVB flow. The formation of
large bubbles in the wake of knots may further perturb the structure
of the outer disk wind.
 
Large-scale flares over similar long timescales, followed by a
re-arrangement of outflow density and velocity distribution, were
recently observed in long-term numerical simulations of an extended
disk wind interacting with the stellar magnetosphere
\citep{Fendt2009}. This scenario could be promising to explain the
DG Tau jet properties, although simulations over larger spatial scales
would be needed to compare with our observations. Alternatively,
a stellar magnetic cycle could be at the origin of the time
variability observed in the launching process or the period of 2.5
years might correspond to an excentric binary companion perturbing the
inner disk at periastron passage. Studies of jet wiggling and
interferometric imaging of DG Tau could help in testing this last
alternative.

\subsubsection{Episodic magnetic bubbles}

Based on recent laboratory experiments of radiatively cooled plasma
jets, \citet{Ciardi2009} propose yet another possible ejection
mechanism for the DG Tau jet. Starting from a highly wound-up field
configuration with $|B_\phi| \propto r^{-1} \gg B_p$ and a strong
axial current, two outflow components are generated: a magnetic
bubble (or cavity) accelerated by gradients of the magnetic pressure
and surrounded by a shell of swept-up ambient material, and a
magnetically confined narrow jet on the bubble axis.  This process is
akin to the magnetic tower modeled numerically e.g. by
\citet{Kato2004}, although it is followed over longer timescales and
with higher resolution.  Repetitive outflow cavities are produced as
the plasma refills in the launch region. The new cavities are confined
by the matter left over from previous episodes, so that collimation
properties gradually become independent of initial conditions. The
morphology is suggestive of that of DG~Tau, with a
collimation angle of $\le 10\degr$ for the high-density clumps along
the axis. However, the bubbles already inflate very close to the disk
so that the conical geometry of the MVB flow is not well
reproduced. Additional confinement of the bubbles by a surrounding
disk wind may be required to explain this property.  Further work in
this direction would be extremely valuable, as would numerical
estimates of the required $|B_\phi|$ in the DG Tau disk.

\section{Conclusions}
\label{sec:conclusions}
\begin{itemize}

\item We observed an onion-like kinematic structure in the blueshifted
jet with a conical medium-velocity (MVB) flow at V $\simeq$ -100 \kms\
surrounding a more tightly collimated high-velocity (HVB) beam at
V $\simeq$ -200 \kms. The velocity fluctuations and electronic density
distribution along the jet show little variation between 1998 and 2005. However,
little \feii\ emission is observed from the low-velocity (LV) flow
component at $\simeq -30,-40$ \kms\ reported in optical lines at 30 AU
from the jet axis.
 
\item By comparing the intensities in \feii\ and \oi\ in the DG Tau
jet at two epochs, we infer evidence for iron depletion by a factor 10
at speeds below 100 \kms, and by a factor 3 in the high-velocity
range. Simultaneous data are required to confirm this result.  Mass-flux rates for the blueshifted jet lobe inferred from three independent methods are
$1.6(\pm 0.8 \times 10^{-8} M_{\odot} yr^{-1}$ in each of
the MVB and HVB ranges, representing 0.02--0.2 of the disk accretion
rate each.

\item The counterjet is dominated by a striking bubble structure.  A similar faint bubble structure is also detected in the blueshifted lobe. The radial and tangential
velocity in the red lobe is lower than in the blue lobe by a factor $\simeq$
0.7. The observed central obscuration, if caused by extinction by the
disk, favors the similarity solution for the DG~Tau disk surface density
distributions derived by \citet{Isella2010} rather than a sharply
truncated power law.
    
\item The two new knots present in the blueshifted lobe indicate  the presence of
a variability timescale of 2.5 years over the last 10 years,  two times
shorter than estimated by \citet{Pyo2003}, and 140 times longer than
the stellar rotation period. The current bubble structures appear to
have formed in the wake of the  high-velocity knot observed by \citet{Pyo2003},
ejected in 1999. Observations more closely spaced in time would help
to clarify the knot/bubble dynamics and the origin of these quasi
periodic eruptions. Searches for periodic changes in mass accretion
rate and for jet wiggling would also constrain the process responsible
for these transient events.

\item The velocities of the \feii\  MVB flow are too high to be
accounted for by disk photo-evaporation models.  Origin of the MVB component in
entrainment of an outer photo-evaporated disk wind seems also unlikely owing to the derived
narrow HVB jet widths. 
The strong iron depletion also
rules out an origin in a stellar wind, or in sideways ejected matter
from the (3 times less depleted) HVB beam.  In addition, neither the
classical X-wind model \citep{Shang1998,Cai2008} nor the conical wind
found in recent simulations of stellar magnetosphere-disk interaction
\citep{Romanova2009} are able to reproduce the kinematics and opening
angles in the \feii\ MVB. A dusty MHD disk wind appears to be the
most plausible origin for this component.

\item The extended MHD disk-wind solution proposed to fit optical
rotation signatures in DG Tau \citep{Pesenti2004} compares more
favorably with observations. It predicts launch radii of 0.08--0.25 AU
and 0.25--1.5 AU, respectively, for the HVB and MVB velocity ranges,
consistent with the higher iron depletion at lower speed. The
predicted mass-fluxes also agree with observed values if the
accretion rate was relatively high in DG Tau over 2002--2004, $ \simeq
5 \times 10^{-7} M_{\odot} yr^{-1}$.  However, contribution from
magnetospheric reconnexions or magnetic towers may not be negligible,
modifying the predicted flow structure. Spectro-imaging monitoring of
accretion and ejection signatures in DG Tau over a few years are
essential to elucidate the ejection processes at work in this
prototypical source.
 
\end{itemize}

\begin{acknowledgements}
      We acknowledge discussions with Jonathan Ferreira and   
     we are grateful to Fabio De Colle for providing the functional
     form used for the density fit in Eq.~\ref{eq:ne}, and to Deirdre
     Coffey for providing the transverse PV diagrams in \oi\ and \sii\ used in
     Fig.~\ref{fig:pvtrans}.  Vanessa Agra-Amboage and Sylvie Cabrit
     wish to acknowledge financial and travel support through the
     Marie Curie Research Training Network JETSET (Jet simulations,
     Experiments and Theory) under contract MRTN-CT-2004-005592. We also acknowledge financial support from the Programme National de Physique Stellaire. Finally, we thank the referee for his/her report.
\end{acknowledgements}

\bibliographystyle{aa}
\bibliography{biblio}

\end{document}